\documentclass[letter,oneside,9pt,twocolumn,article]{aiaa}
\usepackage{graphicx}
\usepackage{rotating}
\usepackage{bm}
\usepackage[bf,normalsize]{subfigure}
\usepackage{adjustbox}
\usepackage{tabularx}
\usepackage{multirow}

\newcommand{\color}[1]{}

\renewcommand{\vec}[1]{\bm{#1}}

\author{  B.\ Parent\thanks{Associate Professor, Aerospace and Mechanical Engineering,  bparent@arizona.edu}~,~~P.\ Thoguluva Rajendran\thanks{Graduate Student, Aerospace and Mechanical Engineering}~,~and~A.\ Omprakas\thanks{Graduate Student, Aerospace and Mechanical Engineering}\\
          {\it University of Arizona, Tucson, AZ 85721}\\
                    }

\title{Electron Losses in Hypersonic Flows\footnote{This article will appear soon in Physics of Fluids.  Copyright 2021 Bernard Parent, Prasanna Thoguluva Rajendran, Ajjay Omprakas. This article is distributed under a Creative Commons Attribution (CC BY) License.}}

\abstract{ 
The first comprehensive study of electron gains and losses in hypersonic air flows including the full coupling between the non-neutral plasma sheaths and the quasi-neutral plasma flow is here presented. Such is made possible by the use of advanced numerical methods that overcome the stiffness associated with the plasma sheaths. The coupling between the sheaths, the electron temperature in non-equilibrium, and the ambipolar diffusion within the quasi-neutral plasma flow is found to be critical to predict accurately electron losses and, thus, plasma density around hypersonic vehicles. This is because   electron cooling coming from the non-neutral sheaths affects significantly electron temperature everywhere in the plasma and, therefore, the  electron-temperature-dependent loss processes of ambipolar diffusion and dissociative recombination. Results obtained show that electron loss to the surface due to catalyticity dominates over electron loss within the plasma due to dissociative recombination either (i) at high altitude where the dynamic pressure is low, or (ii) at low Mach number, or (iii) when the vehicle has a sharp leading edge. 
}

\nomenclature{
  \begin{nomenclaturelist}{Roman symbols}
   \item[$A$] coefficient used to determine reaction rates in Arrhenius form

  \end{nomenclaturelist}

}

\begin{document}
\maketitle
%\makenomenclature

\section{Introduction}

\dropword Plasma electron density {\color{red}around a hypersonic vehicle is a result of the balance between  electron gains and electron losses.} Electron gains can originate from chemical reactions such as Townsend ionization or associative ionization or from {\color{red}thermionic emission} from the surface. Electron losses can originate from chemical reactions within the flow, from chemical reactions taking place at the surface through catalysis, or from convection out of the domain through the outflow boundary. 

\begin{table*}[!t]
\scalefont{0.92}
\begin{center}
\begin{threeparttable}
\tablecaption{Adjusted Dunn-Kang model.}
\begin{tabular}{cccccccccc} 
\toprule
~&~&\multicolumn{4}{c}{Forward reaction} & \multicolumn{4}{c}{Backward reaction} \\
 \cmidrule(lr){3-6}\cmidrule(lr){7-10}
\multicolumn{2}{c}{Reaction} & $T$ & $A$ & $n$ & $E$ & $T$  & $A$ & $n$ & $E$ \\ 
\midrule
1 & $\rm O_2 + M_1 \rightleftarrows 2O+M_1$ &$T$ & $3.6 \cdot 10^{18}$  & $-1$    & $59,500 \, R$  
                                          &$T$  & $3.0 \cdot 10^{15}$  & $-0.5$  & 0\\
2 & $\rm N_2 + M_2 \rightleftarrows 2N+M_2$ &$T$  & $1.9 \cdot 10^{17}$ & $-0.5$ & $113,000 \, R$ 
                                          &$T$  & $1.1 \cdot 10^{16}$  & $-0.5$  & 0\\
3 & $\rm NO + M_3 \rightleftarrows N+O+M_3$ &$T$  & $3.9 \cdot 10^{20}$ & $-1.5$ & $75,500\, R$ 
                                          &$T$ & $1.0 \cdot 10^{20}$  & $-1.5$  & 0\\
4 & $\rm O + NO \rightleftarrows N+O_2$ &$T$ & $3.2 \cdot 10^{9}$ & 1 & $19,700 \, R$ 
                                          &$T$ & $1.3 \cdot 10^{10}$  & $1.0$  & $3,580 \, R$\\
5 & $\rm O + N_2 \rightleftarrows N+NO$ &$T$ & $7 \cdot 10^{13}$ & 0 & $38,000\, R$ 
                                          &$T$ & $1.56 \cdot 10^{13}$  & $0$  & 0\\
6 & $\rm N + N_2 \rightleftarrows 2N+N$ &$T$ & $4.085 \cdot 10^{22}$ & $-1.5$ & $113,000\, R$ 
                                          &$T$ & $2.27 \cdot 10^{21}$  & $-1.5$  & 0\\
                                          
7 & $\rm N + O \rightleftarrows NO^+ + e^- $ & $T$  & 5.3 $\cdot$ 10$^{12}$  & 0.0 & $32000 \, R$ & $T_{\rm e}$ & $5.87 \cdot 10^{17}$ & $-0.2998$ & $100 \, R$ \\

8 & $\rm O + e^- \rightleftarrows O^++2e^-$ &$T_{\rm e}$ & 6.37 $\cdot$ 10$^{16}$ & $0.0029$ & $477,190\, R$ 
                                          &$T_{\rm e}$ & $2.2 \cdot 10^{40}$  & $-4.5$  & 0\\
9 & $\rm N + e^- \rightleftarrows N^++2e^-$ &$T_{\rm e}$ & 1.06 $\cdot$ 10$^{18}$ & $-0.2072$ & $629,700\, R$ 
                                          &$T_{\rm e}$ & $2.2 \cdot 10^{40}$  & $-4.5$  & 0\\
10 & $\rm O_2 + e^- \rightleftarrows O_2^++2e^-$ &$T_{\rm e}$ & 2.33 $\cdot$ 10$^{16}$ & $0.1166$ & $567,360\, R$ 
                                          &$T_{\rm e}$ & $2.2 \cdot 10^{40}$  & $-4.5$  & 0\\
11 & $\rm N_2 + e^- \rightleftarrows N_2^++2e^-$ &$T_{\rm e}$ & 1.58 $\cdot$ 10$^{16}$ & $0.1420$ & $536,330\, R$ 
                                          &$T_{\rm e}$ & $2.2 \cdot 10^{40}$  & $-4.5$  & 0\\
12 & $\rm NO + e^- \rightleftarrows NO^++2e^-$ &$T_{\rm e}$ & 5.63 $\cdot$ 10$^{18}$ & $-0.2607$ & $686,030\, R$ 
                                          &$T_{\rm e}$ & $2.2 \cdot 10^{40}$  & $-4.5$  & 0\\

13 & $\rm O + O \rightleftarrows O_2^+ + e^- $ & $T$  & 1.1 $\cdot$ 10$^{13}$  & 0 & $81200 \, R$ & $T_{\rm e}$ & $1.52 \cdot 10^{18}$ & $-0.3411$ & $881\, R$\\

14 & $\rm O + O_2^+ \rightleftarrows O_2+O^+$ &$T$ & $2.92 \cdot 10^{18}$ & $-1.11$ & $28,000\, R$ 
                                          &$T$ & $7.8 \cdot 10^{11}$  & $0.5$  & 0\\
15 & $\rm N_2 + N^+ \rightleftarrows N+N_2^{+}$ &$T$ & $2.02 \cdot 10^{11}$ & 0.81 & $13,000 \, R$ 
                                          &$T$ & $7.8 \cdot 10^{11}$  & $0.5$  & 0\\

16 & $\rm N + N \rightleftarrows N_2^+ + e^- $ & $T$ & 2.0 $\cdot$ 10$^{13}$  & 0 & $67700 \, R$ & $T_{\rm e}$ &  $4.65 \cdot 10^{17}$ & $-0.2493$ & $7\, R$ \\

17 & $\rm O_2 + N_2 \rightleftarrows NO + NO^+ + e^-$ &$T$ & $1.38 \cdot 10^{20}$ & $-1.84$ & $141,000\, R$ 
                                         &$T_{\rm e}$ & $1.0 \cdot 10^{24}$  & $-2.5$  & 0\\
18 & $\rm NO + N_2 \rightleftarrows N_2 + NO^+ + e^-$ &$T$ & $2.2 \cdot 10^{15}$ & $-0.35$ & $108,000\, R$ 
                                          &$T_{\rm e}$ & $2.2 \cdot 10^{26}$  & $-2.5$  & 0\\
19 & $\rm O + NO^+ \rightleftarrows NO + O^+$ &$T$ & $3.63 \cdot 10^{15}$ & $-0.6$ & $50,800\, R$ 
                                          &$T$ & $1.5 \cdot 10^{13}$  & $0$  & 0\\
20 & $\rm N_2 + O^+ \rightleftarrows O + N_2^{+}$ &$T$ & $3.4 \times 10^{19}$ & $-2$ & $23,000\, R$ 
                                          &$T$ & $2.48 \cdot 10^{19}$  & $-2.2$  & 0\\
21 & $\rm N + NO^+ \rightleftarrows NO + N^+$ &$T$ & $1 \cdot 10^{19}$ & $-0.93$ & $61,000\, R$ 
                                          &$T$ & $4.8 \cdot 10^{14}$  & $0$  & 0\\
22 & $\rm O_2 + NO^+ \rightleftarrows NO + O_2^{+}$ &$T$ & $1.8 \cdot 10^{15}$ & 0.17 & $33,000\, R$ 
                                          &$T$ & $1.8 \cdot 10^{13}$  & $0.5$  & 0\\
23 & $\rm O + NO^+ \rightleftarrows O_2 + N^+$ &$T$ & $1.34 \cdot 10^{13}$ & 0.31 & $77,270\, R$ 
                                          &$T$ & $1.0 \cdot 10^{14}$  & $0$  & 0\\
24 & $\rm NO + O_2 \rightleftarrows NO^+ + e^- + O_2$ &$T$ & $8.8 \cdot 10^{15}$ & $-0.35$ & $108,000\, R$ 
                                          &$T_{\rm e}$ & $8.8 \cdot 10^{26}$  & $-2.5$  & 0\\
25 & $\rm O_2 + O \rightleftarrows 2O + O$ &$T$ & $9 \cdot 10^{19}$ & $-1$ & $59,500\, R$ 
                                          &$T$ & $7.5 \cdot 10^{16}$  & $-0.5$  & 0\\
26 & $\rm O_2 + O_2 \rightleftarrows 2O + O_2$ &$T$ & $3.24 \cdot 10^{19}$ & $-1$ & $59,500\, R$ 
                                          &$T$ & $2.7 \cdot 10^{16}$  & $-0.5$  & 0\\
27 & $\rm O_2 + N_2 \rightleftarrows 2O + N_2$ &$T$ & $7.2 \cdot 10^{18}$ & $-1$ & $59,500\, R$ 
                                          &$T$ & $6.0 \cdot 10^{15}$  & $-0.5$  & 0\\
28 & $\rm N_2 + N_2 \rightleftarrows 2N + N_2$ &$T$ & $4.7 \cdot 10^{17}$ & $-0.5$ & $113,000\, R$ 
                                          &$T$ & $2.72 \cdot 10^{16}$  & $-0.5$  & 0\\
29 & $\rm NO + M_4 \rightleftarrows N + O + M_4$ &$T$ & $7.8 \cdot 10^{20}$ & $-1.5$ & $75,500\, R$ 
                                          &$T$ & $2.0 \cdot 10^{20}$  & $-1.5$  & 0\\
\bottomrule
\end{tabular}
\begin{tablenotes}
\item[{a}] The universal gas constant $R$ must be set to 1.9872	cal/K$\cdot$mol. $A$ has units of $\textrm{cm}^3\cdot(\textrm{mole}\cdot \textrm{s})^{-1}\cdot \textrm{K}^{-n}$ or of $\textrm{cm}^6\cdot(\textrm{mole}\cdot \textrm{s})^{-1}\cdot \textrm{K}^{-n}$. $E$ has units of cal/mole. The rate is given by $A T^n \exp(-E/RT).$
\item[{b}] $\rm M_1=N,~NO$;~ $\rm M_2=O,~NO,~O_2$;~ $\rm M_3=O_2,~N_2$;~ $\rm M_4=O,~N,~NO$.
\item[{c}] The forward rates for 7, 13, and 16 are taken from \cite{pf:2007:boyd}. The forward rate coefficients for reactions 8, 9, 10, 11, 12 are found using BOLSIG+ \cite{psst:2005:hagelaar,pcpp:1992:morgan}. The backward rates for 7 and 13 are found using BOLSIG+ \cite{psst:2005:hagelaar,jgr:1974:walls}. The backward rate for reaction 16 is found using BOLSIG+ \cite{psst:2005:hagelaar,jap:2021:abdoulanziz}. Other reactions are taken from Dunn-Kang \cite{nasa:1973:dunn}.
\end{tablenotes}
\label{tab:parentdunn}
\end{threeparttable}
\end{center}
\end{table*}

Determining which electron loss mechanism is the most important  for a particular hypersonic vehicle geometry and flight conditions can be challenging. Performing experiments to assess plasma density at hypersonic speeds poses significant challenges due to the large flow enthalpy in the Mach number range where plasma forms (Mach 10 and higher). Not only performing experiments at these high enthalpies is a challenge but measuring properties in such flows poses additional difficulties. Perhaps for these reasons, experimental measurements of electron density in hypersonic flows are scarce  and limited to a few measurements at select locations (eg.\ RAM-C-II and OREX). No experiment so far has yielded how much of the electron loss is due to surface catalyticity and how much is due to ion-electron recombination within the plasma bulk.

Numerical simulations of electron loss within hypersonic plasma flows with all the relevant physics included are also particularly difficult to obtain. To yield an accurate prediction of electron loss through surface catalyticity the flow solver needs to predict ambipolar diffusion accurately because surface catalyticity depends on electron flux to the surface with the latter being a function of ambipolar diffusion. But because ambipolar diffusion scales with {\color{red}$(1+T_{\rm e}/T)$} with $T$ the translational temperature of the heavy particules and $T_{\rm e}$ the temperature of the electrons, an accurate prediction of electron diffusion can only be achieved if the electron temperature is well modeled. Few numerical simulations to date have modeled electron temperature accurately. The vast majority of previous studies did not incorporate an electron energy transport equation and assumed that the electron temperature is simply equal to the vibrational temperature (see for instance Refs.\ \cite{jsr:2008:keidar,jtht:2009:tchuen,aiaa:2009:wan,aiaapaper:2014:surzhikov,rjpcb:2015:surzhikov,ijhmt:2021:kim,aiaapaper:2021:sawicki}). A few prior studies which did include an electron energy transport \cite{jtht:1991:candler,jtht:2013:farbar} were missing important physics related to non-neutral plasma sheath effects and thus yielded a potentially incorrect prediction of electron temperature. 

Why is incorporating non-neutral effects within  the electron temperature in non-equilibrium  particularly difficult?  {\color{red}The difficulty originates from the large discrepancy between the time scales of the plasma sheath (sub nanosecond) and the one of the plasma bulk (microsecond and higher). Implicit integration techniques used in hypersonic CFD could overcome this stiffness if all the physical phenomena associated with the smallest time scales could be written as diffusion or source terms.  When simulating plasma sheaths, this is not possible unfortunately because of the \emph{drift} motion of the charged species with respect to the bulk. The charged species drift  is a convection phenomenon with the wave speed being proportional to the electric field. There is thus a very important difference between a non-neutral plasma mixture and a neutral plasma mixture. Indeed, for a neutral plasma mixture, the motion of each component with respect to the bulk can be expressed fully using diffusion terms (see species diffusion rates in a multicomponent gas in Ref.\ \cite{book:1999:giovangigli} for instance). For a non-neutral plasma mixture including plasma sheaths, the motion of the electrons and ion species with respect to the bulk can not be expressed using diffusion terms only and require the use of first derivatives to account for the drift. Because the stiffness originating from the drift terms (i.e., the convection terms) can not be overcome using block-implicit methods, this leads to an excessive amount of computing to reach convergence, thus forcing the use of too-coarse meshes thus tainting the results with excessive amounts of numerical error.  }

Although some previous studies have claimed to incorporate the effect of the sheath in the simulation of hypersonic flows (see \cite{pst:2016:zhiwei,pop:2011:sotnikov,iccem:2019:cheng,optik:2018:cong}), such used the word sheath loosely to refer to the quasi-neutral plasma layer. If sheath is defined as usual as the non-neutral region near the surface then only one numerical study (outlined in Ref.\ \cite{pf:2021:parent}) has incorporated so far the effect of the sheath in the simulation of hypersonic flows. This was made possible through the use of newly developed algorithms \cite{jcp:2013:parent,jcp:2014:parent,aiaa:2016:parent} that overcome the stiffness associated with the simulation of plasma sheaths. The study showed  that, for a wedge with a sharp leading edge, the sheath had a profound effect on electron temperature not only in the vicinity of the surface but also deep within the quasi-neutral region. Such  was attributed to (i) the amount of electron cooling taking place within the sheath being commensurate with the electron heating elsewhere and (ii) the high electron thermal conductivity spreading the sheath cooling effect to the rest of the plasma. Thus, because ambipolar diffusion depends on electron temperature, and because electron temperature everywhere in the plasma is affected by the sheath, it is critical to incorporate the sheath effects for an accurate simulation of electron loss by diffusion to the surface.

In this paper, we will use the advanced numerical methods outlined in \cite{jcp:2013:parent,jcp:2014:parent,aiaa:2016:parent} that permit to simulate the full coupling between the sheath and the plasma bulk. Using these novel methods we will assess under what conditions electron loss by catalyticity dominates over electron loss through chemical reactions within the plasma bulk. This will be done over a range of flight conditions relevant to hypersonic flight through parametric studies of the flight altitude, flight Mach number, and vehicle size and geometry.

\begin{table*}[!t]
\scalefont{0.92}
\begin{center}
\begin{threeparttable}
\tablecaption{Adjusted Park model.}
\begin{tabular}{cccccccccc} 
\toprule
~&~&\multicolumn{4}{c}{Forward reaction} & \multicolumn{4}{c}{Backward reaction} \\
 \cmidrule(lr){3-6}\cmidrule(lr){7-10}
\multicolumn{2}{c}{Reaction} & $T$ & $A$ & $n$ & $E$ & $T$  & $A$ & $n$ & $E$ \\ 
\midrule
1 & $\rm N_2 + M_1 \rightleftarrows N + N + M_1$ & $T$   & 3.0 $\cdot$ 10$^{22}$  & $-1.6$ & $113200 \, R$  & $T$ &\multicolumn{3}{c}{Equilibrium constant} \\

2 & $\rm N_2 + M_2 \rightleftarrows N + N + M_2$ & $T$  & 7.0 $\cdot$ 10$^{21}$  & $-1.6$ & $113200 \, R$ & $T$ & \multicolumn{3}{c}{Equilibrium constant}  \\

3 & $\rm N_2 + e^- \rightleftarrows N + N + e^-$ & $T_{\rm e}$ & 3.0 $\cdot$ 10$^{24}$  & $-1.6$ & $113200 \, R$ & $T_{\rm e}$ & \multicolumn{3}{c}{Equilibrium constant} \\

4 & $\rm O_2 + M_1 \rightleftarrows O + O + M_1$ & $T$  & 1.0 $\cdot$ 10$^{22}$  & $-1.5$ & $59500 \, R$ & $T$ & \multicolumn{3}{c}{Equilibrium constant} \\

5 & $\rm O_2 + M_2 \rightleftarrows O + O + M_2$ & $T$  & 2.0 $\cdot$ 10$^{21}$  & $-1.5$ & $59500 \, R$ & $T$ & \multicolumn{3}{c}{Equilibrium constant} \\

6 & $\rm NO + M_3 \rightleftarrows N + O + M_3$ & $T$  & 1.1 $\cdot$ 10$^{17}$  & 0.0 & $75500 \, R$ & $T$ & \multicolumn{3}{c}{Equilibrium constant} \\

7 & $\rm NO + M_4 \rightleftarrows N + O + M_4$ & $T$  & 5.0 $\cdot$ 10$^{15}$  & 0.0 & $75500 \, R$ & $T$ & \multicolumn{3}{c}{Equilibrium constant}  \\

8 & $\rm NO + O \rightleftarrows N + O_2 $ & $T$  & 8.4 $\cdot$ 10$^{12}$  & 0.0 & $19400 \, R$ & $T$ & \multicolumn{3}{c}{Equilibrium constant}  \\

9 & $\rm N_2 + O \rightleftarrows NO + N $ & $T$  & 5.7 $\cdot$ 10$^{12}$  & $0.42$ & $42938 \, R$ & $T$ & \multicolumn{3}{c}{Equilibrium constant} \\

10 & $\rm N + O \rightleftarrows NO^+ + e^- $ & $T$  & 5.3 $\cdot$ 10$^{12}$  & 0.0 & $32000 \, R$ & $T_{\rm e}$ & $5.87 \cdot 10^{17}$ & $-0.2998$ & $100 \, R$ \\

11 & $\rm O + O \rightleftarrows O_2^+ + e^- $ & $T$  & 1.1 $\cdot$ 10$^{13}$  & 0 & $81200 \, R$ & $T_{\rm e}$ & $1.52 \cdot 10^{18}$ & $-0.3411$ & $881\, R$\\

12 & $\rm N + N \rightleftarrows N_2^+ + e^- $ & $T$ & 2.0 $\cdot$ 10$^{13}$  & 0 & $67700 \, R$ & $T_{\rm e}$ &  $4.65 \cdot 10^{17}$ & $-0.2493$ & $7\, R$ \\

13 & $\rm NO^+ + O \rightleftarrows N^+ + O_2 $ & $T$   & 1.0 $\cdot$ 10$^{12}$  & 0.5 & $77200 \, R$ & $T$& \multicolumn{3}{c}{Equilibrium constant}\\

14 & $\rm N^+ + N_2 \rightleftarrows N_2^+ + N $ & $T$   & 1.0 $\cdot$ 10$^{12}$  & 0.5 & $12200 \, R$ & $T$ & \multicolumn{3}{c}{Equilibrium constant}\\

15 & $\rm O_2^+ + N \rightleftarrows N^+ + O_2 $ & $T$   & 8.7 $\cdot$ 10$^{13}$  & 0.14 & $28600 \, R$ & $T$ & \multicolumn{3}{c}{Equilibrium constant}\\

16 & $\rm O^+ + NO \rightleftarrows N^+ + O_2 $ & $T$  & 1.4 $\cdot$ 10$^{5}$  & 1.90 & $26600 \, R$ & $T$  & \multicolumn{3}{c}{Equilibrium constant}\\

17 & $\rm O_2^+ + N_2 \rightleftarrows N_2^+ + O_2 $ & $T$   & 9.9 $\cdot$ 10$^{12}$  & 0.00 & $40700 \, R$ & $T$ & \multicolumn{3}{c}{Equilibrium constant}\\

18 & $\rm O_2^+ + O \rightleftarrows O^+ + O_2 $ & $T$   & 4.0 $\cdot$ 10$^{12}$  & $-0.09$ & $18000 \, R$ & $T$ & \multicolumn{3}{c}{Equilibrium constant}\\

19 & $\rm NO^+ + N \rightleftarrows O^+ + N_2 $ & $T$   & 3.4 $\cdot$ 10$^{13}$  & $-1.08$ & $12800 \, R$ & $T$ & \multicolumn{3}{c}{Equilibrium constant}\\

20 & $\rm NO^+ + O_2 \rightleftarrows O_2^+ + NO $ & $T$   & 2.4 $\cdot$ 10$^{13}$  & 0.41 & $32600 \, R$ & $T$ & \multicolumn{3}{c}{Equilibrium constant}\\

21 & $\rm NO^+ + O \rightleftarrows O_2^+ + N $ & $T$   & 7.2 $\cdot$ 10$^{12}$  & 0.29 & $48600 \, R$ & $T$ & \multicolumn{3}{c}{Equilibrium constant}\\

22 & $\rm O^+ + N_2 \rightleftarrows N_2^+ + O $ & $T$  & 9.1 $\cdot$ 10$^{11}$  & 0.36 & $22800 \, R$ & $T$  & \multicolumn{3}{c}{Equilibrium constant}\\

23 & $\rm NO^+ + N \rightleftarrows N_2^+ + O $  & $T$  & 7.2 $\cdot$ 10$^{13}$  & 0.00 & $35500 \, R$ & $T$ & \multicolumn{3}{c}{Equilibrium constant}\\

24 & $\rm O^+ + e^- \rightarrow O + {\it hv} $ & $T_{\rm e}$  & 1.07 $\cdot$ 10$^{11}$  & $-0.52$ & $0$ &  \multicolumn{4}{c}{None}\\
25 & $\rm N^+ + e^- \rightarrow N + {\it hv} $ & $T_{\rm e}$  & 1.52 $\cdot$ 10$^{11}$  & $-0.48$ & $0$ &  \multicolumn{4}{c}{None}\\

26 & $\rm O + e^- \rightleftarrows O^+ + e^- + e^- $ & $T_{\rm e}$  & 6.37 $\cdot$ 10$^{16}$  & $0.0029$ & $477190 \, R$ & $T_{\rm e}$ & $2.2\cdot 10^{40}$ & $-4.5$ & 0\\

27 & $\rm N + e^- \rightleftarrows N^+ + e^- + e^- $ & $T_{\rm e}$  & 1.06 $\cdot$ 10$^{18}$  & $-0.2072$ & $629700 \, R$ & $T_{\rm e}$ & $2.2\cdot 10^{40}$ & $-4.5$ & 0\\

28 & $\rm O_2 + e^- \rightleftarrows O_2^+ + e^- + e^-$ &$T_{\rm e}$ & 2.33 $\cdot$ 10$^{16}$ & $0.1166$ & $567360\, R$ 
                                          &$T_{\rm e}$ & $2.2 \cdot 10^{40}$  & $-4.5$  & 0\\
29 & $\rm N_2 + e^- \rightleftarrows N_2^+ + e^- + e^-$ &$T_{\rm e}$ & 1.58 $\cdot$ 10$^{16}$ & $0.1420$ & $536330\, R$ 
                                          &$T_{\rm e}$ & $2.2 \cdot 10^{40}$  & $-4.5$  & 0\\
30 & $\rm NO + e^- \rightleftarrows NO^+ + e^- + e^-$ &$T_{\rm e}$ & 5.63 $\cdot$ 10$^{18}$ & $-0.2607$ & $686030\, R$ 
                                          &$T_{\rm e}$ & $2.2 \cdot 10^{40}$  & $-4.5$  & 0\\
\bottomrule
\end{tabular}
\begin{tablenotes}
\item[{a}] The rate coefficient is given by $A T^n \exp(-E/RT)$; The universal gas constant $R$ must be set to 1.9872	cal/K$\cdot$mol. $A$ has units of $\textrm{cm}^3\cdot(\textrm{mole}\cdot \textrm{s})^{-1}\cdot \textrm{K}^{-n}$ or of $\textrm{cm}^6\cdot(\textrm{mole}\cdot \textrm{s})^{-1}\cdot \textrm{K}^{-n}$. $E$ has units of cal/mole. 
\item[{b}] $\rm M_1=N,~O,~N^+,~O^+$; $\rm M_2=N_2,~O_2,~NO,~N_2^+,~O_2^+,~NO^+$; $\rm M_3= N,~O,~NO,~N^+,~O^+$; $\rm M_4=N_2,O_2,N_2^+,~O_2^+,~NO^+$.
\item[{c}] The forward rates for 10, 11, and 12 are taken from \cite{pf:2007:boyd}. The backward rates for 10, 11 are found using BOLSIG+ \cite{psst:2005:hagelaar,jgr:1974:walls}. The backward rates for reaction 12 is found using BOLSIG+ \cite{psst:2005:hagelaar,jap:2021:abdoulanziz}. The backward rates for 26--30 are taken from \cite{nasa:1973:dunn}. Reaction 8 is taken from \cite{jcp:1997:bose} and reaction 9 from \cite{jcp:1996:bose}. The forward rates for reactions 26--30 are found using BOLSIG+ \cite{psst:2005:hagelaar,pcpp:1992:morgan}. Reactions 24 and 25 are taken from \cite{jtht:1993:park}. Other reactions are taken from \cite{book:1990:park}.
\end{tablenotes}
\label{tab:parentpark}
\end{threeparttable}
\end{center}
\end{table*}

\section{Physical Model}

{\color{red}
The momentum equation for the bulk of the mixture (with the ``bulk'' here referring to the mixture of neutrals and charged species) is based on the Navier-Stokes equations with source terms to account for the force the electric field exerts on a non-neutral plasma. Either charged or neutral, each species  has a different velocity than the bulk mixture velocity and its motion is obtained through the solution of a separate mass conservation transport equation. For the neutrals, the velocity difference with respect to the bulk is set proportional to the product between the mass fraction gradient and the mass diffusion coefficient with the latter obtained from the Lennard-Jones potentials \cite{nasa:1962:svehla}. For the electrons and ions, the velocity difference with respect to the bulk involves both a drift and a diffusion component (the so-called ``drift-diffusion'' model). The drift velocity corresponds to the electric field multiplied by the sign of the charge and by the mobility of the species in question, while the diffusion coefficient is obtained from the mobility through the Einstein-Smoluchowski relation. Expressions for the mobilities used here can be found in \cite{pf:2021:parent}. The mobilities take into consideration collisions with the neutrals but neglect collisions with other charged species. This is well justified for the problems solved herein because the plasma is \emph{weakly-ionized} (i.e., the ionization fraction is less than 0.1\% or so). Indeed, for a weakly-ionized plasma the forces originating from viscous terms or from collisions with other charged species  are much smaller than forces originating from collisions between charged species and neutrals (see \cite{jcp:2011:parent} for a discussion).  } 

The vibrational temperature of nitrogen and the electron temperature are obtained through  separate transport equations as outlined in \cite{jpp:2007:parent,pf:2021:parent}.  The gas is assumed thermally perfect but calorically non-perfect with the enthalpies for each species determined from the NASA Glenn high-temperature polynomials \cite{nasa:2002:mcbride}. The translational, rotational, and electronic temperatures are assumed equal for all heavy species and set to the bulk gas temperature. Only the $\rm N_2$ vibrational temperature and the electron temperature differ from the bulk gas temperature. The vibrational temperature at the wall is assumed equal to the wall temperature {\color{red}while the electron temperature at the surface is extrapolated from the nearby boundary node because electrons travel within the sheath towards the surface.} The thermal conductivity for the mixture is obtained from the Mason and Saxena relation, and the viscosity of the mixture is obtained from Wilke's mixing rule. The electric field components that are needed to determine the charged species velocities are determined from an electric field potential equation based on Gauss's law.  A complete description of all transport equations making up the physical model can be found in Ref.\ \cite{pf:2021:parent}.

The species production and destruction rates are determined through an eleven-species air chemical solver. The species include $\rm e^-$, $\rm N_2$, $\rm O_2$, $\rm N$, $\rm O$, $\rm NO$, $\rm NO^+$, $\rm O^+$, $\rm N^+$, $\rm N_2^+$, $\rm O_2^+$. Because the reaction rates for hypersonic airflow are not fully understood and may be prone to error, we will first compare the results obtained with several chemical solvers to available experimental data. Then, we will choose the best performing chemical solver to do the parametric studies.

One chemical solver we will use is the one by Dunn-Kang \cite{nasa:1973:dunn}. Although many reaction rates within the Dunn-Kang model are well known to have considerable error, the model as a whole performs surprisingly well in predicting ionization and surface heat flux in hypersonic flows. Perhaps this is due to several reaction rates not being well known at the time the model was created and having been adjusted (or guessed) in order to obtain good agreement with experimental data. 

A second model we will use is an ``adjusted Dunn-Kang'' model where all the reactions are the same as the Dunn-Kang model except for the following modifications done to the rates of reactions that involve electrons: (i) the reaction rates for associative ionization are taken from \cite{pf:2007:boyd}; (ii) the Townsend ionization (electron impact ionization) reactions of all species are found using BOLSIG+ \cite{psst:2005:hagelaar} with the cross-sectional data obtained from Ref.\ \cite{pcpp:1992:morgan}; (iii) the dissociative recombination rates are obtained from BOLSIG+ with the cross-sectional data obtained from Refs.\ \cite{jgr:1974:walls} and \cite{jap:2021:abdoulanziz}. The complete list of reactions making up the adjusted Dunn-Kang model is outlined in Table \ref{tab:parentdunn}.

A third model we will use in this study is the so-called ``Park  model'' \cite{book:1990:park}. Such is now commonly understood to refer to the set of reactions proposed by Park in 1990 but with the following changes: (i) the reaction rates for associative ionization are taken from Boyd \cite{pf:2007:boyd}; (ii) the reaction rates for $\rm NO+O \leftrightarrows N + O_2$ and $\rm NO+N \leftrightarrows N_2 + O$ are taken from Bose et al.\ in Refs.\ \cite{jcp:1997:bose} and \cite{jcp:1996:bose} respectively. Also, as suggested by Park, the controlling temperature for the dissociation reactions is $\sqrt{T T_{\rm v}}$, while the controlling temperature for the dissociative recombination reactions and of the reaction $\rm N_2+e^- \rightarrow N+N+e^-$ is  $\sqrt{T_{\rm v} T_{\rm e}}$, and the controlling temperature of the $\rm N+N + e^- \rightarrow N_2 + e^-$ reaction is $\sqrt{T T_{\rm e}}$. Further, when needed, the backward rates are found through the equilibrium constant approach using the NASA Glenn polynomials \cite{nasa:2002:mcbride}.

A fourth model used herein consists of the ``adjusted Park model''. Such is based on the so-called Park model outlined in the previous paragraph but with several modifications to the rates of the reactions that involve electrons: (i) the dissociative recombination reactions are found using BOLSIG+ \cite{psst:2005:hagelaar} with the cross sections taken from Refs.\ \cite{jgr:1974:walls} and \cite{jap:2021:abdoulanziz}; (ii) the three-body recombination reaction rates are taken from Dunn-Kang \cite{nasa:1973:dunn}; (iii) the electron impact ionization reactions are found using BOLSIG+ with the cross sections taken from \cite{pcpp:1992:morgan}. Also, we make a modification to the Park controlling temperatures. Instead of taking the geometric average between the two most important temperatures as suggested by Park, we rather use simply the dominant temperature disregarding the other. As will be demonstrated through some validation cases, this yields a better agreement with flight test data overall. All the reactions comprising the adjusted Park model are shown in Table \ref{tab:parentpark}.

\section{Numerical Methods}

Obtaining the steady-state solution through iterative methods to the physical model listed in the above section poses difficulties. One difficulty lies with the different physical processes within the model having time scales that differ greatly from each other. Indeed, the time scales associated with the motion of the neutrals may be in the order of a few microseconds, but the time scales associated with some chemical reactions or with the motion of the electrons can be several orders of magnitude less. This discrepancy of the time scales is one reason why the  system of equations is stiff and requires a large number of iterations to converge. Such can be overcome  through the use of a block-implicit method. Amongst the various block implicit schemes we have tested, the Diagonally-Dominant Alternate-Direction-Implicit (DDADI) scheme \cite{aiaaconf:1987:bardina,cf:2001:maccormack} was the most successful in overcoming the stiffness originating from the disparate time scales and is here the chosen iterative method for all fluid transport equations. Thus, the mass, momentum, and energy equations of all species (electrons, ions, and neutrals) are integrated in pseudotime in coupled form through the block-implicit DDADI approach.

The disparate time scales are not the only source of stiffness within our system of equations, however. A second source of stiffness comes from some terms within the Gauss-based potential equation amplifying the error associated with charged species densities (see \cite{jcp:2013:parent} for explanations). This source of stiffness becomes particularly problematic when quasi-neutral plasma regions of significant size form within the domain, as is the case for the problems here considered. Because the stiffness associated with the error amplification can not be relieved through block implicit methods, we here follow the approach shown in \cite{jcp:2013:parent,jcp:2014:parent,aiaa:2016:parent} where the stiffness-inducing terms are avoided by obtaining the potential from Ohm's law instead of Gauss's law. To ensure that Gauss's law is satisfied, some source terms are added to the ion transport equations. It is emphasized that such a recast of the equations is done without taking shortcuts or modifying the physical model in any way and is thus strictly a convergence acceleration method. Lastly, fast convergence of the electric field potential equation to steady-state is obtained through a combination of iterative modified approximate factorization (IMAF) \cite{cf:2001:maccormack} and successive over relaxation (SOR). {\color{red}By using the latter methods, convergence to steady-state of the physical model used herein (i.e., including the drift-diffusion model for the charged species and the electric field potential equation based on Gauss's law) requires only about 2 to 3 times more computing than  converging a ``standard'' neutral set of transport equations for hypersonic non-equilibrium flows (i.e., excluding the drift-diffusion model for the charged species and the electric field potential equation).
}

The convection derivatives are discretized through the Roe scheme turned second-order accurate through the MUSCL approach and  the  Van Leer Total Variation Diminishing (TVD) limiter. To prevent carbuncles while not introducing excessive dissipation within the boundary layer, the eigenvalues within  the discretized flux are adjusted through the entropy correction method based on the Peclet number outlined in \cite{aiaa:2017:parent}. Further, to ensure that the densities, pressures, and temperatures do not become negative, the positivity-preserving filter outlined in \cite{aiaaconf:2019:parent} is applied.

\section{Code Validation}

The physical model consisting of the coupling between the Navier-Stokes equations for the neutrals and the drift-diffusion model for the charged species was implemented within the code CFDWARP, an open-source CFD code for {\color{red}ionized and reacting compressible flows}. CFDWARP has been previously validated for  hypersonic plasma flows \cite{jpp:2007:parent,pf:2021:parent}, but  additional validation cases are here presented using the four alternative chemical solvers outlined in the Physical Model section above. We will do so for two well-known sets of hypersonic flight test data: RAM-C-II \cite{nasa:1970:grantham,nasa:1972:jones} and OREX \cite{asvpaper:1995:inouye}.  

\begin{figure}[h]
     \centering
     \subfigure[61 km]{\includegraphics[width=0.35\textwidth]{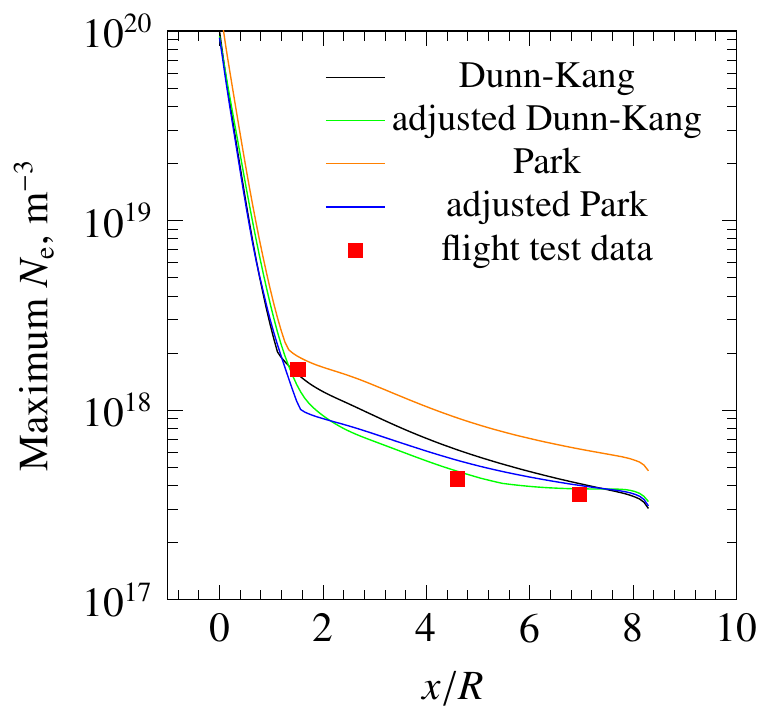}}
     \subfigure[71 km]{\includegraphics[width=0.35\textwidth]{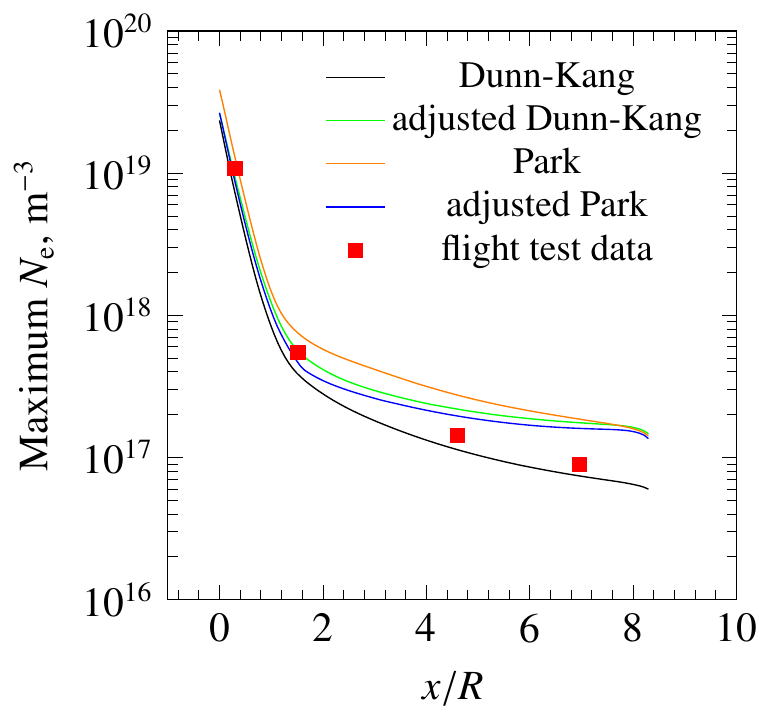}}
     \figurecaption{Comparison between numerical results and RAM-C-II flight test data on the basis of maximum electron number density at (a) 61 km altitude and (b) 71 km altitude.}
     \label{fig:RAMCII_comparison1}
\end{figure}

\begin{figure}[h]
     \centering
     \subfigure[61 km ]{\includegraphics[width=0.32\textwidth]{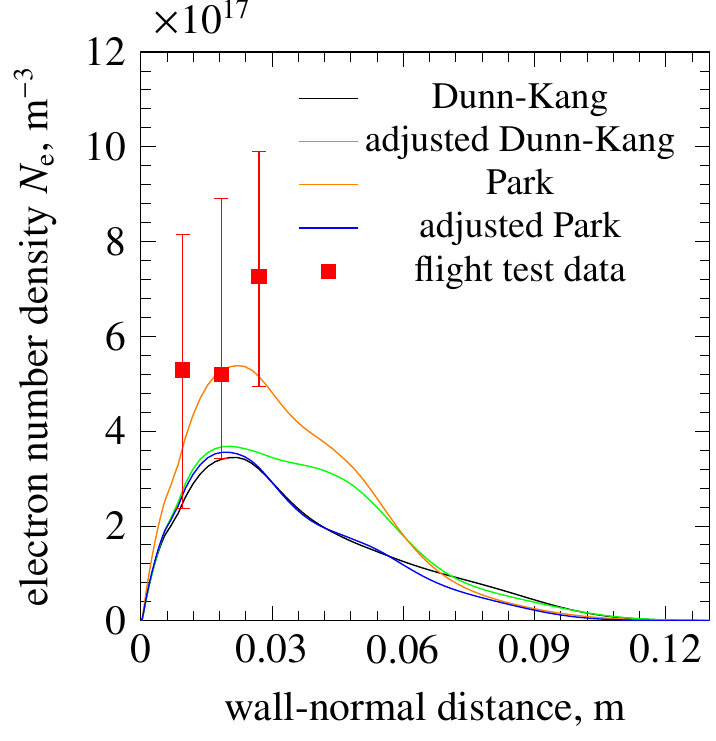}}
     \subfigure[71 km ]{\includegraphics[width=0.32\textwidth]{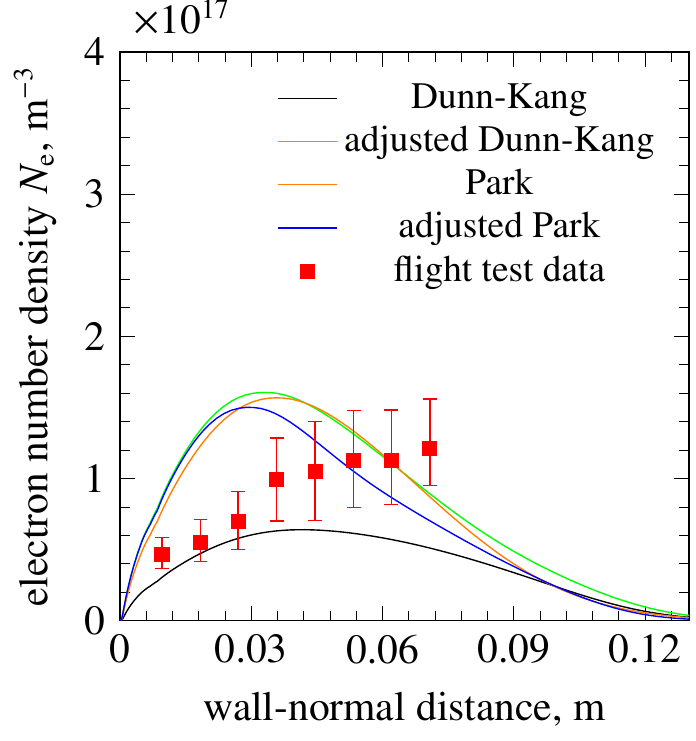}}
     \figurecaption{Comparison between CFDWARP and RAM-C-II flight test data on the basis of electron number density  at $x/R=8.1$ at (a) 61 km altitude and (b) 71 km altitude.}
     \label{fig:RAMCII_comparison2}
\end{figure}

\subsection{RAM-C-II}

 The RAM-C-II vehicle  has a blunt-wedge shaped geometry with a nose-radius $R=0.1524$ metres, a half-cone angle of 9 degrees, and a body-length 8.5 times its nose radius. Microwave reflectometers stationed along the body were used to measure the maximum number density at given streamwise stations. Flight-test data from two altitudes of 61 and 71 km are compared here with the results from the numerical simulation.  The freestream properties at the 61 km altitude correspond to a Mach number of 23.9, a temperature of 255.9~K, and a dynamic pressure 8~kPa, while the freestream properties at the 71 km altitude correspond to a Mach number of 25.9, a temperature of 217.9~K, and a dynamic pressure of 2.28~kPa. The RAM-C-II data is good to assess electron loss due to the chemical reactions within the flow as opposed to electron loss to the surface. Indeed, for the two altitudes considered, the 2-body dissociative recombination reactions and associative ionization reactions are the primary sources of electron destruction and production, respectively.

 In Fig.\ \ref{fig:RAMCII_comparison1}, the peak electron number densities are compared, at various $x/R$ stations beginning from the nose of the vehicle, with the numerical solution obtained using the four competing chemical models. It can be seen that the adjusted Dunn-Kang and Park models generally perform as well or significantly better than the original models. At an altitude of 61 km, both the adjusted Park and adjusted Dunn-Kang models match experimental data within a few percent either near the leading edge or further downstream. 
 
 Further comparisons are shown in Fig.\ \ref{fig:RAMCII_comparison2} where the electron number density distribution along the wall-normal direction at the location of the electrostatic probe rake is compared. The agreement between the CFD results and the flight test data is not as good in this case. All chemical solvers underpredict by a factor of 2 or so the distance from the surface of the peak number density for the 71~km altitude case. The error is not due to the grid being too coarse: grid convergence studies show that the mesh used here is sufficiently fine that no discernible difference in the plots would be seen if using a finer mesh. Rather, it seems likely that the discrepancy originates from experimental error {\color{red}(or from how the electron density is inferred from the experimental data)} because the maximum electron density along the electrostatic probe rake differs from the one obtained using microwave reflectometers by a factor of 2 or so at a similar axial location. 
 
{\color{red} The discrepancy between CFDWARP and the RAM-C-II results along the electrostatic rake is postulated to be due to physical phenomena not well understood at the time of the experiments leading to incorrect inference of electron density.  Indeed, electron density is not directly measured by the electrostatic probes. Rather, the probes measure ion fluxes from which the electron density far from the probe surface is inferred. The relationship between ion flux to the probe and nearby electron density was obtained through experiments performed on the ground at different flow conditions than experienced in flight. It is plausible that the much higher enthalpy of the flow in the flight test leads to a different relationship than expected between ion flux to the probe and electron density, leading to significant error in the reported electron density. A detailed 3D numerical simulation of the electrostatic rake interacting with a re-entry boundary layer including all physical processes of importance (such as the full coupling between the plasma sheaths surrounding the probes and the ambipolar diffusion of the electrons and ions towards the probes) would provide more insight into how to infer more accurately electron density from surface ion flux in hypersonic flight tests. Such is beyond the scope of this paper, however, and is left for future work. }

\begin{figure*}[!h]
     \centering
     \subfigure[59.6 km]{\includegraphics[height=0.24\textwidth]{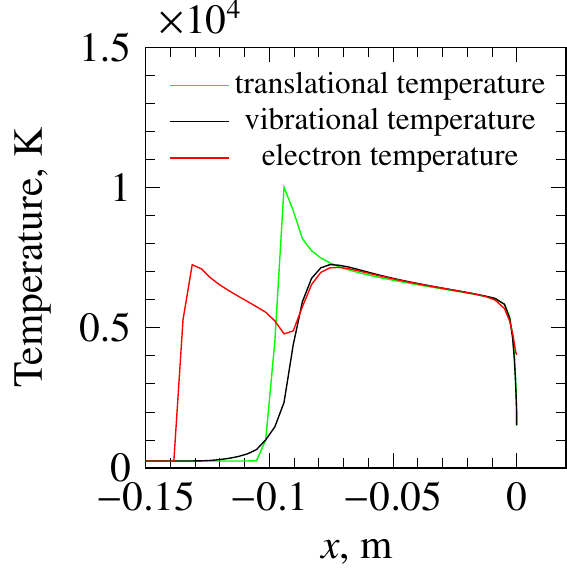}}
     \subfigure[79.9 km]{\includegraphics[height=0.24\textwidth]{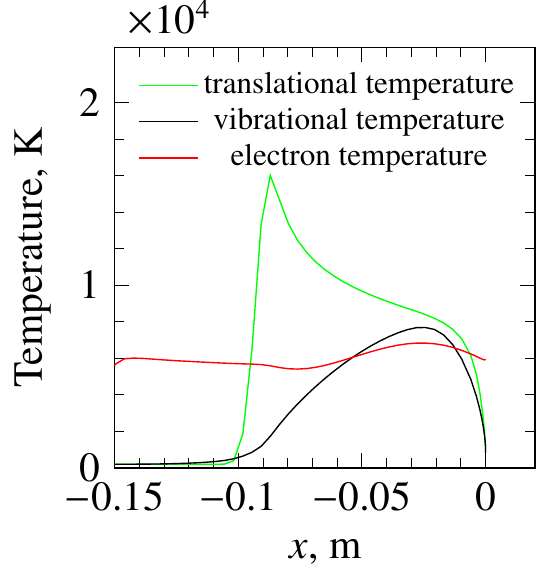}}
     \subfigure[96.8 km]{\includegraphics[height=0.24\textwidth]{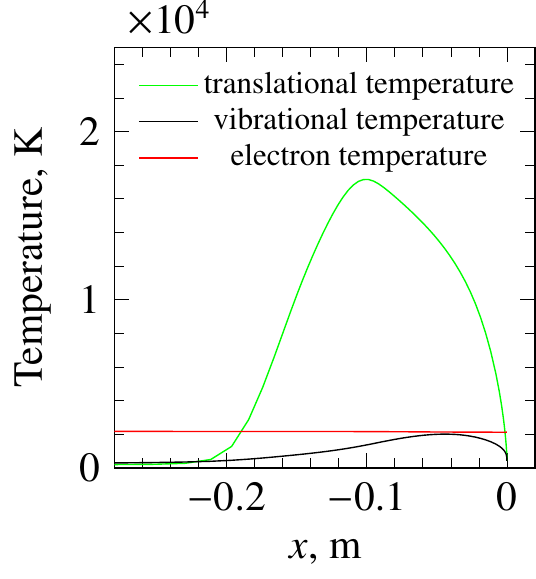}}
     \figurecaption{Comparison of translational, vibrational and electron temperature along stagnation streamline {\color{red}at an altitude of (a) 59.6~km, (b) 79.9~km, and (c) 96.8~km} for the OREX case run using the Dunn-Kang model.}
     \label{fig:OREX_Tcomparison}
\end{figure*}

\begin{figure}[!h]
     \centering
     \subfigure[]{\includegraphics[width=0.31\textwidth]{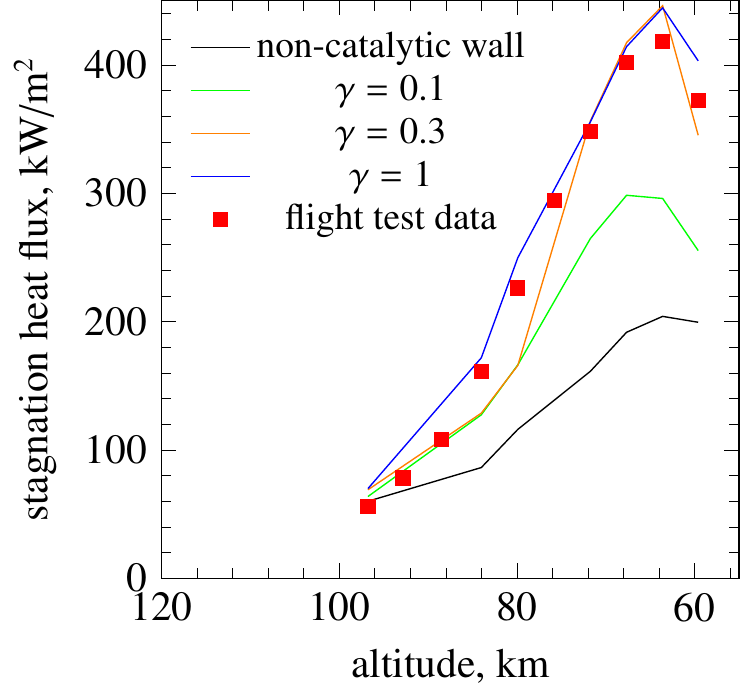}}
     \subfigure[]{\includegraphics[width=0.31\textwidth]{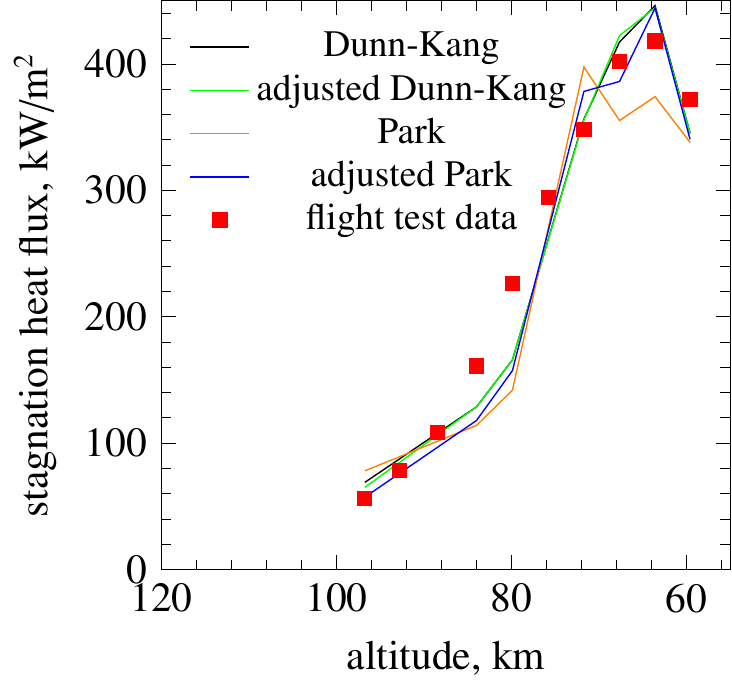}}
     \figurecaption{Comparison between OREX flight test data and CFDWARP results on the basis of stagnation heat flux using  (a) the Dunn-Kang model varying the catalytic recombination coefficient  ($\gamma=\gamma_{\rm O}=\gamma_{\rm N}$) between 0 and 1;  and using (b) different chemical models keeping $\gamma$ fixed to 0.3.}
     \label{fig:OREX_heatflux-CFDWARP}
\end{figure}

\begin{figure}[!t]
     \centering
     \subfigure[84 km]{\includegraphics[width=0.35\textwidth]{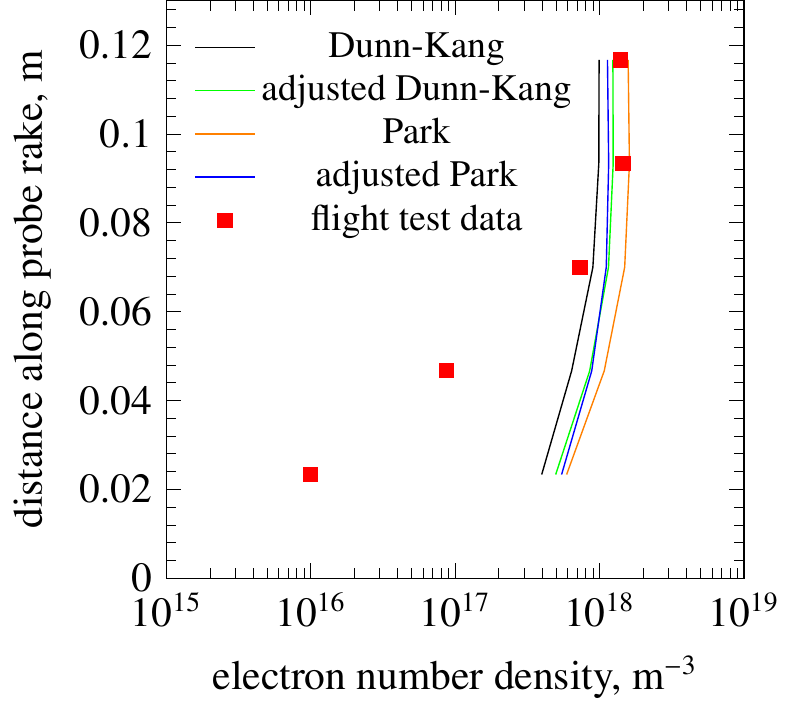}}
     \subfigure[96 km]{\includegraphics[width=0.35\textwidth]{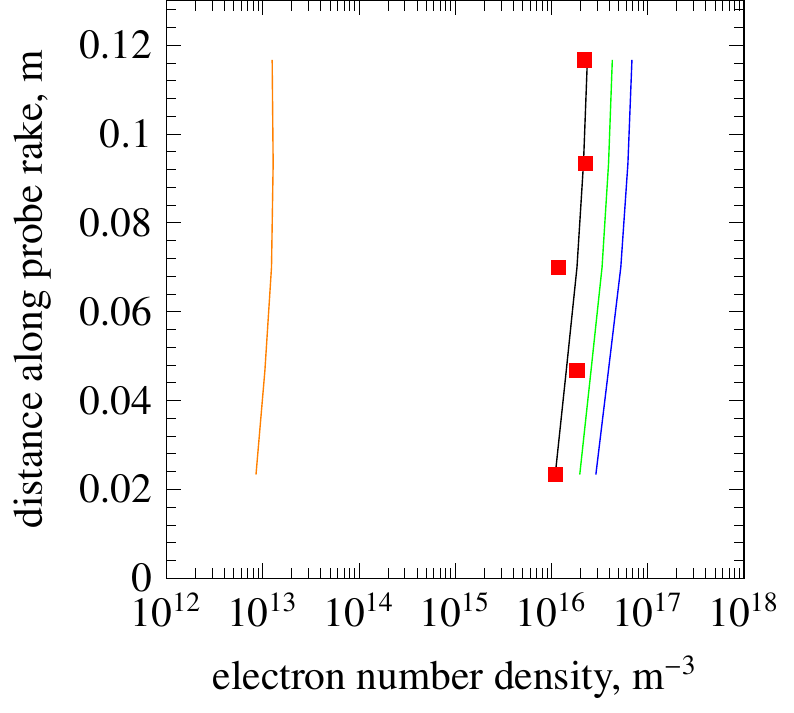}}
     \figurecaption{Comparison between OREX flight test data and numerical results on the basis of electron number density at the electrostatic probe locations at (a) 84 km altitude and (b) 96 km altitude.}
     \label{fig:OREX_Necomparison-CFDWARP}
\end{figure}

\subsection{OREX}

The Orbital Reentry Experiment (OREX) \cite{asvpaper:1995:inouye} was conducted by the National Aerospace Laboratory and the National Space Development agency of Japan in 1994  \cite{sw:2002:doihara}. The OREX vehicle is composed of a 1.35~m radius nose followed by a half-cone wedge angle of 50 degrees. OREX adds to the RAM-C-II experiments through surface heat flux measurements. Such were obtained by monitoring the temperature of the surface at the stagnation point as a function of time. How much heat  the plasma transferred to the surface was determined through the increase of the surface temperature in time.

%For all OREX results obtained with CFDWARP shown below, we will use a mesh size of $173 \times 60$ nodes. A grid convergence study shows that this results in a numerical error less than 15\% for heat flux to the surface or electron densities.

As with the RAM-C-II case, the OREX test case exhibits a large degree of thermal non-equilibrium. This is especially true at high altitude where the electron and vibrational temperatures differ from the translational temperature by almost an order of magnitude (see Fig. \ref{fig:OREX_Tcomparison}). Such has a large impact on various reaction rates involving electrons and, thus, on the electron number density. Further, at the highest altitudes considered, the electron temperature is significantly different from the vibrational temperature. OREX thus serves as a good test bed to validate the capabilities of CFDWARP in predicting electron density and surface heat flux in the presence of significant thermal non-equilibrium.

We first compare CFDWARP results with OREX flight test data on the basis of heat flux to the surface at the stagnation point. We do so over the range of altitudes 60 to 100 kms. As shown in Fig.\ \ref{fig:OREX_heatflux-CFDWARP}, one parameter that plays an important role is the catalytic recombination coefficient for atomic oxygen and atomic nitrogen. When the recombination coefficient is varied between 0 (no catalyticity) to 0.3 (30\% of full catalyticity), the difference between the OREX and CFDWARP results is reduced significantly. Very good  agreement is obtained with flight test data by setting the recombination coefficient to 0.3. {\color{red}Such is the value we will use from now in this subsection to obtain all subsequent OREX results.} Interestingly, as seen in Fig.\ \ref{fig:OREX_heatflux-CFDWARP}, changing the chemical solver has a minimal impact on heat flux at the lower altitudes. Even at the highest altitude considered where thermal and chemical non-equilibrium effects are more important, and where we would expect a more important difference between the different chemical solvers, there is at most a 30\% difference in the heat flux prediction between the four chemical models tested. 
\begin{figure}[t]
     \centering
     \includegraphics[width=0.39\textwidth]{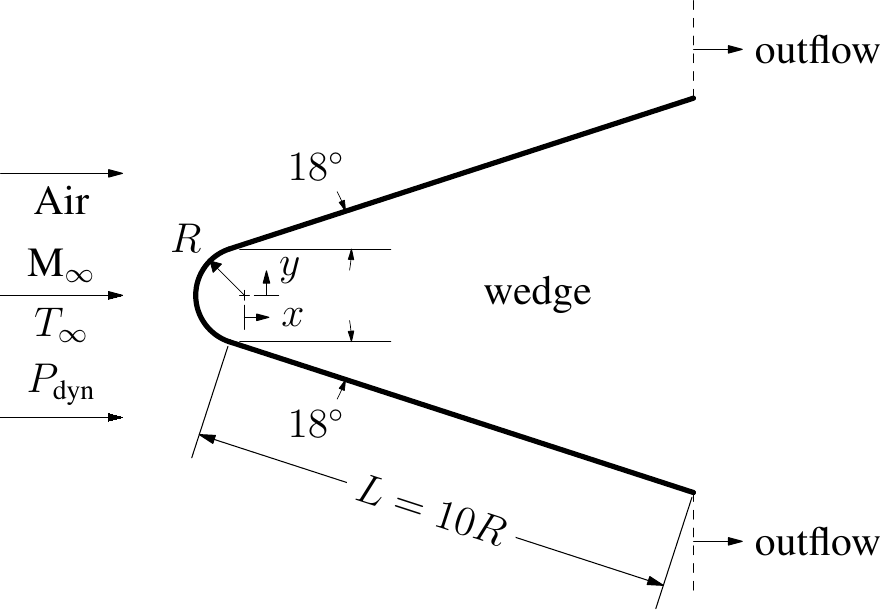}
     \figurecaption{Problem setup}
     \label{fig:schematic}
\end{figure}
\begin{figure}[!t]
    \centering
    \subfigure[$P_{\rm dyn}=3$ kPa]{\includegraphics[width=0.33\textwidth]{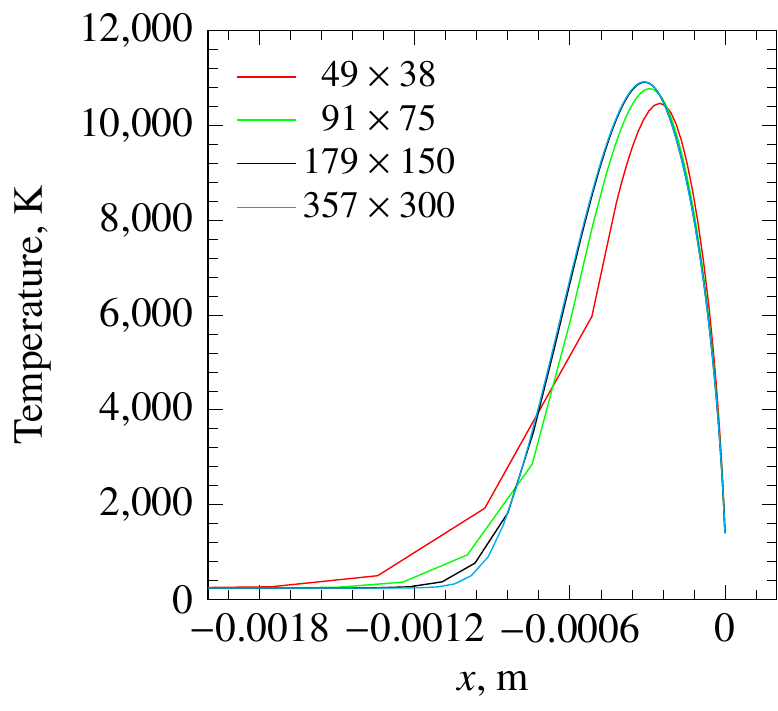}}
    \subfigure[$P_{\rm dyn}=50$ kPa]{\includegraphics[width=0.33\textwidth]{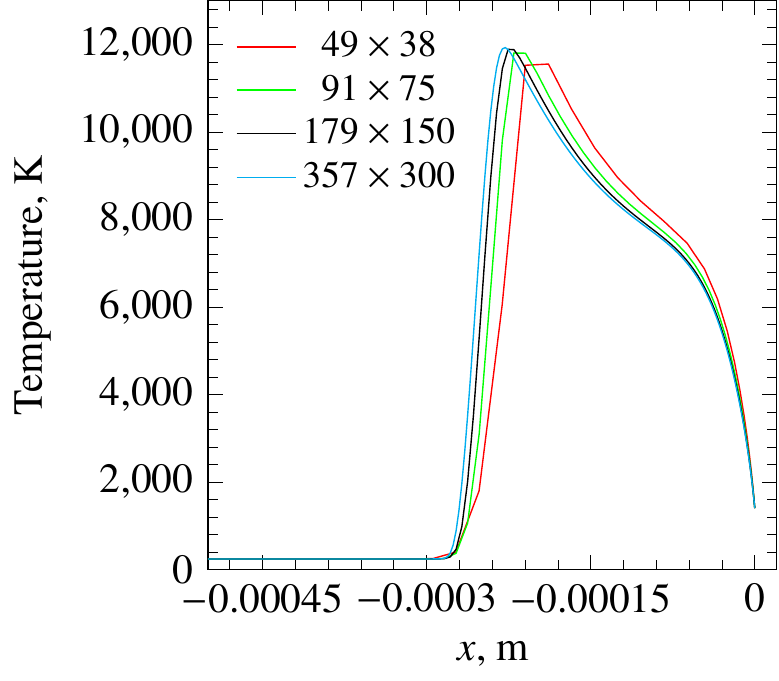}}
    \figurecaption{{\color{red}Effect of grid size on gas temperature on the stagnation streamline for (a)  $P_{\rm dyn}=3$~kPa, (b)  $P_{\rm dyn}=50$~kPa; the freestream Mach number is fixed to 18 and $R$ to 1~mm. }}
    \label{fig:gcs}
\end{figure}
As can be seen in Fig.\ \ref{fig:OREX_Necomparison-CFDWARP}, good agreement between the various chemical solvers  and OREX is obtained overall with respect to electron number density throughout the altitude range 70-100~km. One exception is the non-corrected Park model which underpredicts plasma density at high altitude by three orders of magnitude throughout the electrostatic rake. This is attributed to the Park model using a temperature of $\sqrt{T T_{\rm v}}$ to determine associative ionization. Such a strategy to determine the chemical reactions is prone to yield large error because it gives equal weight to the translational and vibrational temperatures at all times  when finding the rate coefficient. This is prone to error for cases where the rate coefficient is more strongly dependent on one temperature than the other and closer agreement can be obtained by setting the controlling temperature to the dominant temperature. Indeed, simply changing the controlling temperatures from $\sqrt{T T_{\rm v}}$ to $T$ and from $\sqrt{TT_{\rm e}}$ to $T_{\rm e}$ yields results in much close agreement with experimental data as can be seen from the results obtained using the ``Adjusted Park Model''. 

% need to add one plot comparing flight test data and numerical results of surface heat flux

\section{Problem Setup}

The problem we will focus on in this paper consists of a hypersonic flow interacting with a wedge with a rounded leading edge as depicted in Fig.\ \ref{fig:schematic}. Because electron losses will vary significantly for a change in nose radius or freestream conditions, various parametric studies are here performed to gain a better understanding of these physical phenomena over a wide range of conditions. Specifically, the nose radius will be varied between 1~mm and 10~cm, the dynamic pressure between 1 and 50~kPa, {\color{red}and} the Mach number between 12 and 24. Unless otherwise specified, the wall temperature is set to 1400~K.

\begin{figure}[!b]
    \centering
    \subfigure[$P_{\rm dyn}=3$ kPa]{\includegraphics[width=0.307\textwidth]{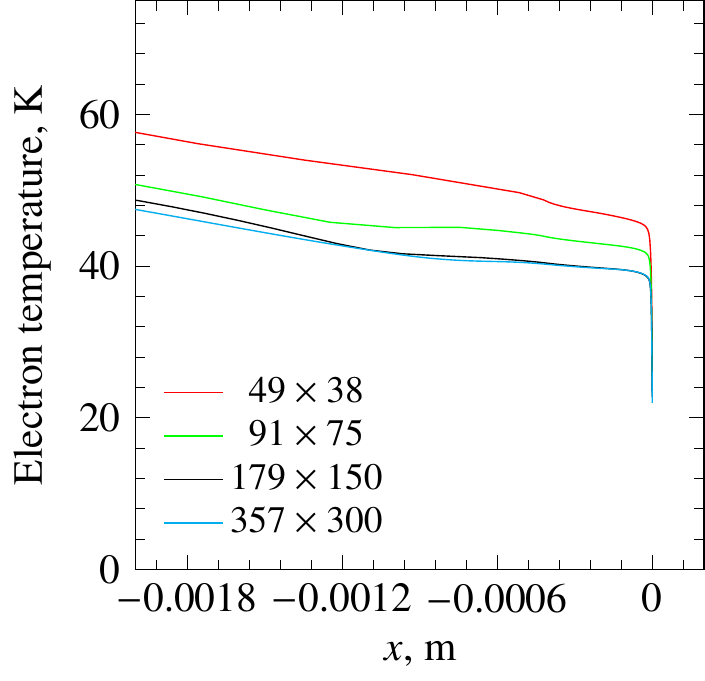}}
    \subfigure[$P_{\rm dyn}=50$ kPa]{\includegraphics[width=0.33\textwidth]{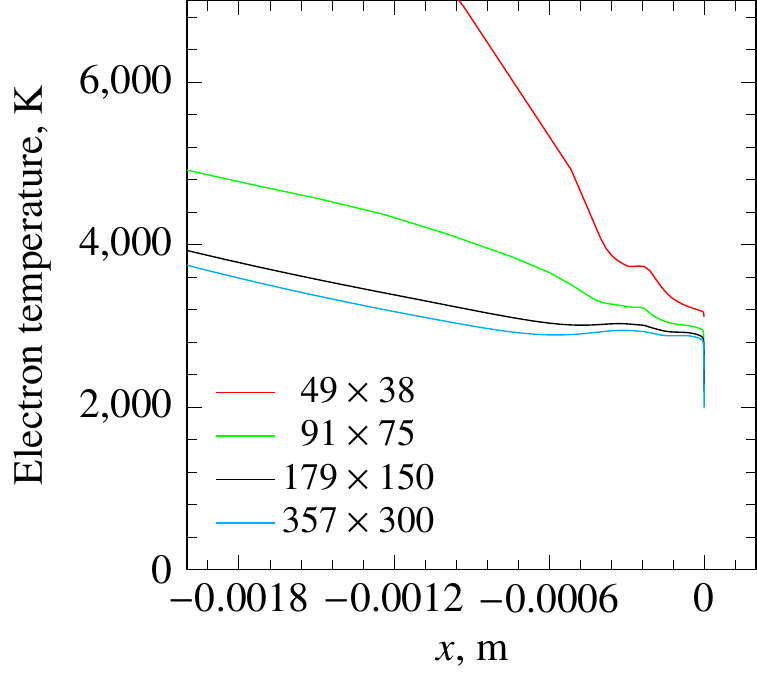}}
    \figurecaption{Effect of grid size on stagnation streamline electron temperature for (a)  $P_{\rm dyn}=3$~kPa, (b)  $P_{\rm dyn}=50$~kPa; the freestream Mach number is fixed to 18 and $R$ to 1~mm. }
    \label{fig:gcs2}
\end{figure}

\begin{figure}[!t]
    \centering
    \subfigure[$P_{\rm dyn}=3$ kPa]{\includegraphics[width=0.32\textwidth]{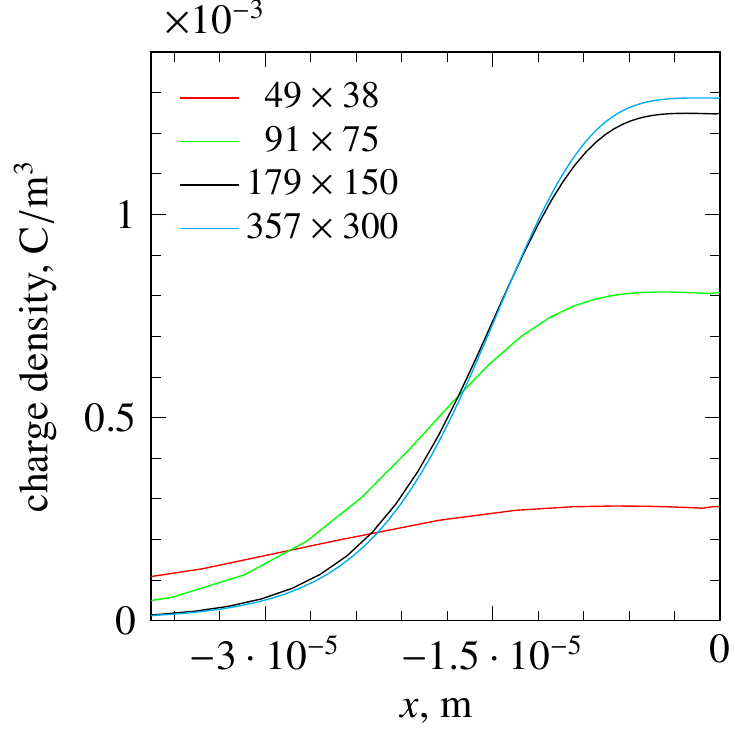}}
    \subfigure[$P_{\rm dyn}=50$ kPa]{\includegraphics[width=0.32\textwidth]{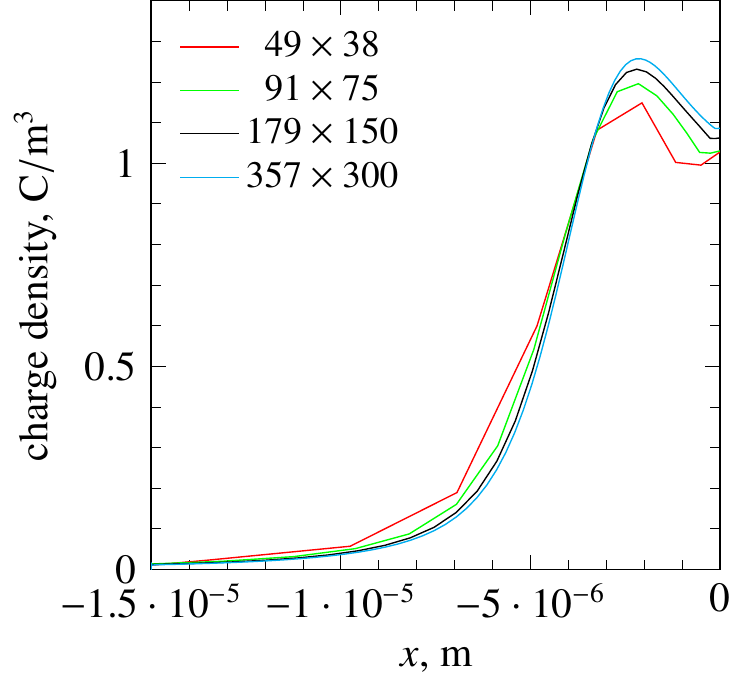}}
    \figurecaption{\color{red}Effect of grid size on net charge density on the stagnation streamline for (a)  $P_{\rm dyn}=3$~kPa, (b)  $P_{\rm dyn}=50$~kPa; the freestream Mach number is fixed to 18 and $R$ to 1~mm. }
    \label{fig:gcs3}
\end{figure}

\begin{figure}[!h]
     \centering
     \includegraphics[width=0.32\textwidth]{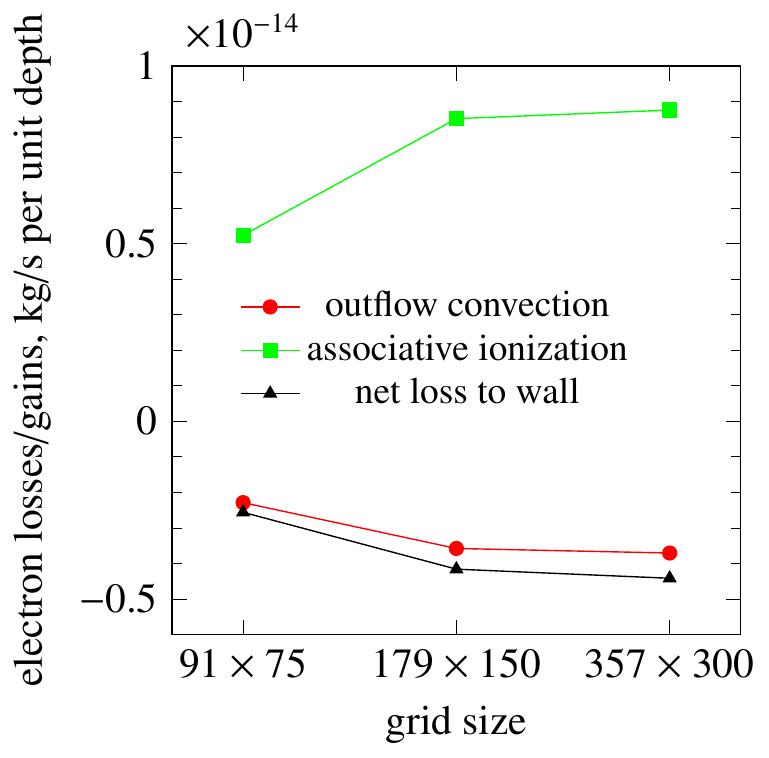}
     \figurecaption{\color{red}Effect of grid size on electron losses and gains integrals (in kg/s per unit depth)  for  $M=18$, $R=1$~mm and $P_{\rm dyn}=3$ kPa.}
     \label{fig:gcs_integrals}
\end{figure}

\section{Numerical Error Assessment}

To determine how fine the grid should be to yield an acceptably small amount of numerical error on the properties of interest, various grid convergence studies were performed. For instance, in  Figs.\ \ref{fig:gcs} and \ref{fig:gcs2}, the effect of the grid density on the electron and gas translational temperatures is shown. {\color{red}Clearly, the error decays more or less as expected for a second order accurate discretization stencil, with minimal differences observed between the two finest meshes. 

Cathode sheaths can sometimes require a hundred grid lines or so to be resolved properly, but this is not the case here. This is because for the problems here considered, the electric field within the sheath is not sufficiently strong to lead to large amounts of Townsend ionization (electron impact ionization). Thus, the coupling between the electric field, the chemical reactions, and the charged species velocities within the sheath is not as strong here as it can be for other problems involving large amounts of Townsend ionization (such as in arc welding for instance). Thus a few tens of grid lines within the sheath is here sufficient to capture accurately the sheath properties and the sheath-induced electron cooling. This is confirmed through grid convergence studies of the net charge density within the sheath in Fig.\ \ref{fig:gcs3}.}

{\color{red}Grid convergence studies were not only performed on local flow properties but also on global parameters (eg.\ the surface integral of electron loss due to catalyticity, the volume integral of electron gains due to associative ionization, etc). The effect of the grid on some surface and volume integrals is shown in Fig.\ \ref{fig:gcs_integrals}.}  The various grid convergence studies show that a grid of $179 \times 150$ results in a numerical error on the local properties less than 5\% and a numerical error on the integrated properties less than 10\%. Unless otherwise specified, we here choose this mesh size to compute all cases. 

\section{Results and Discussion}
\begin{table*}[!t]
\scalefont{0.94}
\centering
\tablecaption{Electron losses and gains (in kg/s per unit depth)  with $M=18$ and $R=5$~mm.}
\begin{tabular}{lcccc}
\toprule
\multirow{2}*{Electron gain/loss mechanism} & \multicolumn{4}{c}{dynamic pressure} \\
\cmidrule{2-5}
& 1 kPa & 3 kPa            & 10 kPa            & 50 kPa              \\
\midrule
Convection through outflow boundary    & $-1.09\times10^{-13}$         & $-8.16\times10^{-12}$ & $-3.59\times10^{-11}$ & $-1.35\times10^{-10}$  \\
Surface catalyticity    & $-1.19\times10^{-13}$     & $-5.62\times10^{-12}$ & $-1.06\times10^{-11}$ & $-2.22\times10^{-11}$   \\
2-body recombination within the plasma   & $-2.70\times10^{-15}$      & $-3.16\times10^{-11}$ & $-9.89\times10^{-10}$ & $-2.32\times10^{-8}$   \\
3-body recombination within the plasma  & $-1.53\times10^{-14}$        & $-6.11\times10^{-13}$ & $-6.20\times10^{-12}$ & $-1.34\times10^{-10}$   \\
Electron impact ionization & 0   & 0 & 0  & 0    \\
Associative ionization & $2.58\times10^{-13}$& $4.68\times10^{-11}$  & $1.04\times10^{-9}$  & $2.35\times10^{-8}$   \\
\bottomrule
\end{tabular}
\label{tab:gainslosses_Pdyn_integrals}
\end{table*}

Several parametric studies are now performed to quantify the importance of the various electron loss mechanisms. We will vary the flight dynamic pressure, the flight Mach number, the leading edge radius, and the wall temperature in ranges that are relevant to hypersonic flight.  Independently of flight dynamic pressure or Mach number, the freestream temperature is always fixed to 240~K, and the wall temperature is always fixed to 1400~K. Only one chemical solver will be used to obtain all the results in this section: the adjusted Park model outlined in Table \ref{tab:parentpark}. We choose this model because, amongst the chemical solvers with up-to-date and accurate reactions for the electron gains and losses, it is the one that was the closest to the OREX and RAM-C-II flight test data.

\subsection{Effect of Dynamic Pressure}

\begin{figure}[!b]
    \centering
    \subfigure[$R=1$~mm]{\includegraphics[width=0.35\textwidth]{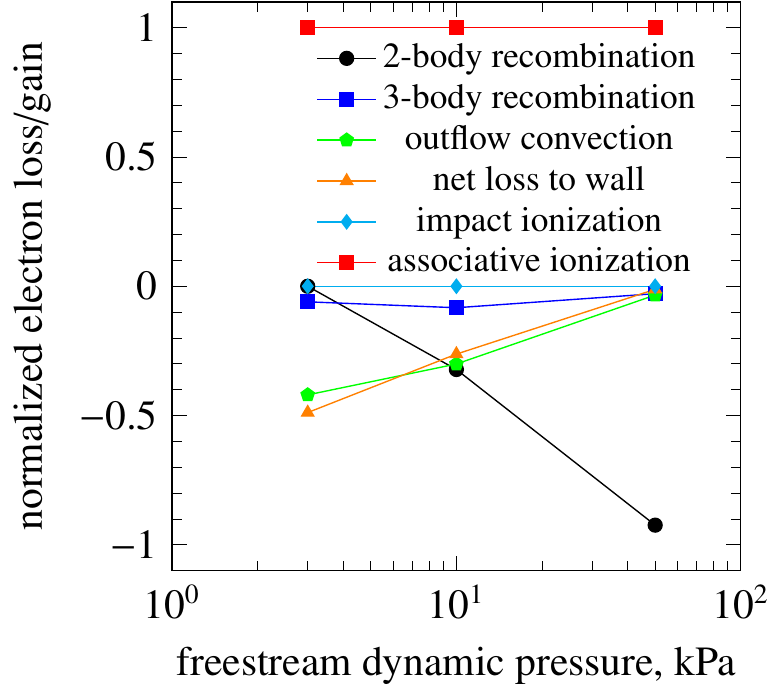}}
    \subfigure[$R=5$~mm]{\includegraphics[width=0.35\textwidth]{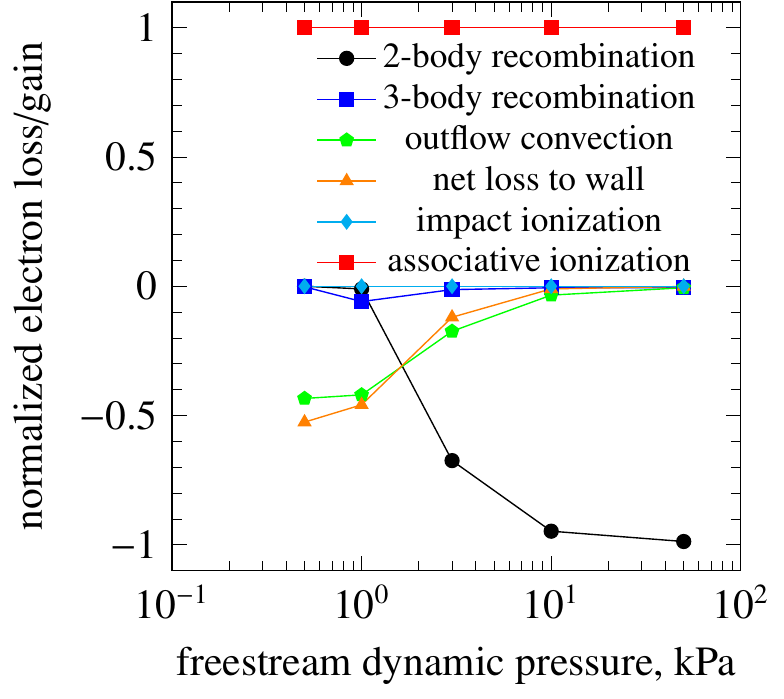}}
    \figurecaption{Impact of flight dynamic pressure on electron gains and losses at Mach 18 for (a) $R=1$~mm, (b) $R=5$~mm.}
    \label{fig:gainslosses_Pdyn}
\end{figure}

We first consider the effect of a change in the flight dynamic pressure. The dynamic pressure is defined following the ``incompressible'' definition as $\frac{1}{2} \rho_\infty q_\infty^2$. Because the freestream temperature is fixed to 240~K, and given the freestream Mach number, the freestream flow speed can thus be readily found from the relationship $q_\infty=M_\infty \sqrt{\gamma R T_\infty}$ with $\gamma$ and $R$ the ratio of the specific heats and the gas constant respectively. Therefore, given the Mach number, changing the dynamic pressure can be understood as a change in the freestream density (or altitude, on which density depends). We nonetheless here prefer to express the parametric study in terms of dynamic pressure rather than density or altitude because most hypersonic flights have a limited range of flight dynamic pressure. This is partly because heat to the surface scales more or less with dynamic pressure and because the aerodynamic lift and drag also scale with dynamic pressure. Thus, in practice, for sufficient lift and not excessive heat flux to the surfaces, hypersonic waveriders typically fly in the dynamic pressure range 1 to 50 kPa. Such is the dynamic pressure range we will investigate on here.

In Table \ref{tab:gainslosses_Pdyn_integrals}, the various electron gains and losses are assessed through surface or volume integrals through the range of dynamic pressure of interest. Electron losses and gains due to chemical reactions within the flow are quantified through volume integrals performed over the entire domain. Electron losses due to surface catalyticity are found through a surface integral of $\rho_{\rm e} \vec{V}^{\rm e} \cdot \vec{n}$ (with $\vec{n}$ the unit normal vector pointing out of the domain, $\vec{V}^{\rm e}$ the electron velocity, and $\rho_{\rm e}$ the electron mass density) along the body of the vehicle. Electron losses due to convection out of the domain are found through a similar surface integral over the outflow boundary. Because all the electron losses and gains are taken into consideration, the sum of all integrals should be zero. This is not exactly the case here because of numerical error: the sum of the integrals is rather about 0.1\% or so of the magnitude of the largest integral. This small error in the sum of the integrals is due to the discretization error and it was verified that such error decreases asymptotically towards zero as the mesh is refined.

For easier comparison between cases, it is convenient to normalize the gains and the losses. We do so by first summing all the gains, and then dividing each gain/loss integral by the sum of the gains. The resulting normalized gains and losses are plotted as a function of the dynamic pressure in Fig.\ \ref{fig:gainslosses_Pdyn} for a baseline flight Mach number of 18. What is immediately apparent from this plot is that the electron gains originate fully from associative ionization and not in any significant portion from electron impact ionization. This is because, with the adjusted Park model used herein, the reaction rate of electron impact ionization remains low and negligible unless the electron temperatures would be in excess of 30000~K. But such is not happening here with the electron temperature remaining well below 10000~K for all cases considered. 

The electron losses are not due to a single process as the gains are, however. As can be seen in Fig.\ \ref{fig:gainslosses_Pdyn}, a variation in dynamic pressure leads to a different  electron loss process dominating. At high dynamic pressure the vast majority of the electrons are lost through 2-body recombination, with few electrons lost through other processes. But at low dynamic pressure, very few electrons are lost through 2-body or 3-body recombination. Rather,  the electron losses  are more or less evenly split between diffusion to the surface and convection out of the outflow boundary.

\begin{figure*}[!t]
    \centering
    \subfigure[$R=5$~mm and $P_{\rm dyn}=1$~kPa]{\includegraphics[width=0.35\textwidth]{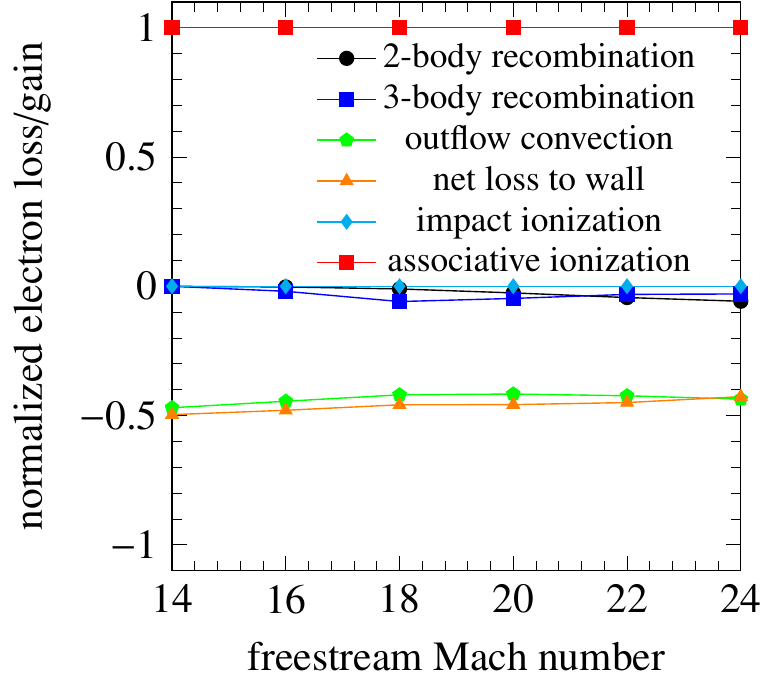}}
    \subfigure[$R=5$~mm and $P_{\rm dyn}=3$~kPa]{\includegraphics[width=0.35\textwidth]{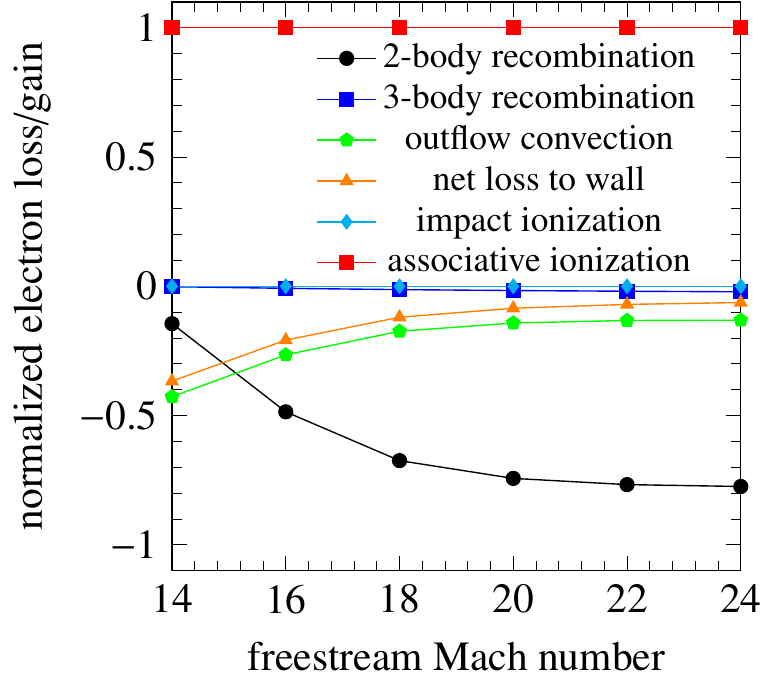}}
    \subfigure[$R=5$~mm and $P_{\rm dyn}=10$~kPa]{\includegraphics[width=0.35\textwidth]{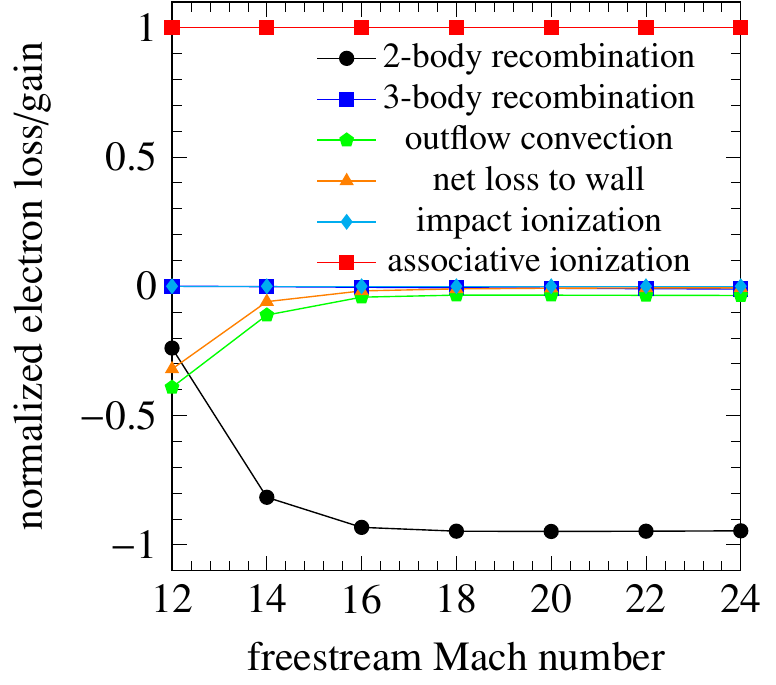}}
    \subfigure[ $R=1$~mm and $P_{\rm dyn}=10$~kPa]{\includegraphics[width=0.35\textwidth]{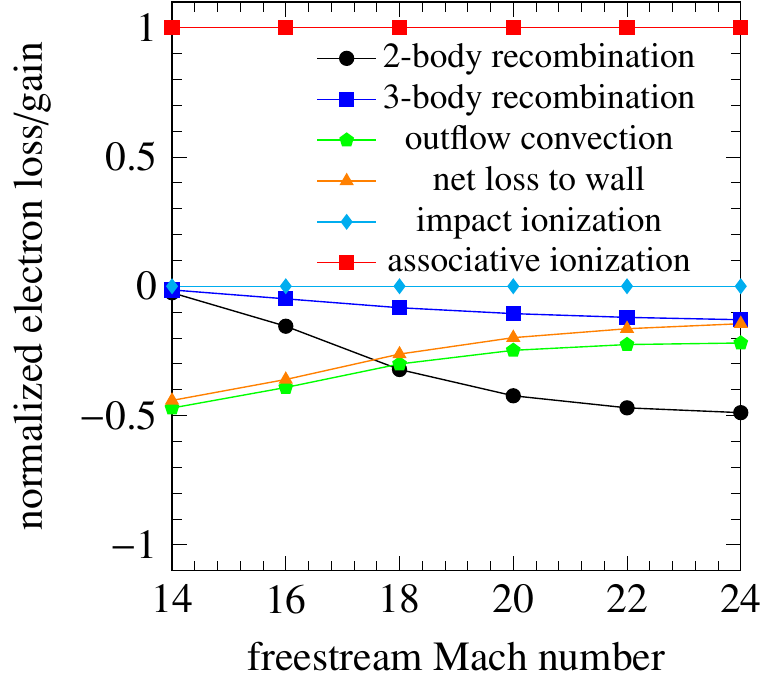}}
    \figurecaption{Impact of flight Mach number on electron gains and losses for (a) $R=5$~mm and $P_{\rm dyn}=1$~kPa, (b) $R=5$~mm and $P_{\rm dyn}=3$~kPa, (c) $R=5$~mm and $P_{\rm dyn}=10$~kPa, and (d) $R=1$~mm and $P_{\rm dyn}=10$~kPa.}
    \label{fig:gainslosses_Mach}
\end{figure*}

The reason for  2-body recombination  losing its importance at lower dynamic pressure is because at a given Mach number (i) a lower dynamic pressure entails a lower flow density and lower charged species densities and (ii) the two-body recombination loss mechanism scales with density of the electrons to the square while the electron loss to the surface is more simply proportional to the density of the electrons. Indeed, at any dynamic pressure here considered, most of the plasma within the domain is quasi-neutral and, therefore, the ion density is about the same as the electron density. Therefore, because the 2-body recombination reaction is proportional to the product between the ion and electron densities, and because the ion density is about the same as the electron density over most of the domain, the reaction rate throughout most of the domain scales more or less with the square of the electron density. Thus, we would expect the ratio between the 2-body recombination losses and the losses at the wall/outflow to vary by about one order of magnitude when the dynamic pressure varies by the same amount. This agrees well with the numerical results observed in Fig.\ \ref{fig:gainslosses_Pdyn}.

\subsection{Effect of Mach Number}

We now proceed to study the effect of the Mach number on the electron gains and losses. We here do so by varying the freestream Mach number between 12 and 24 while keeping the dynamic pressure and vehicle geometry constant. The results are shown in the various plots within Fig.\ \ref{fig:gainslosses_Mach} where each subfigure corresponds to a different combination of flight dynamic pressure and  leading edge radius. In most cases studied, independently of the dynamic pressure or leading edge radius, a similar trend is observed. At low Mach number, the main electron loss mechanisms are surface recombination and outflow convection. But when the Mach number is high, most of the electrons are lost through 2-body recombination (i.e.\ dissociative recombination). 

It may seem peculiar that  2-body recombination as well as 3-body recombination become stronger compared to other loss mechanisms as the Mach number increases. Indeed, we may intuitively expect the recombination processes to be relatively weaker and not stronger as the Mach number increases because (i) an increase of the Mach number at constant dynamic pressure leads to lower flow densities and (ii) the rates of 2- and 3-body recombination scale with the plasma density to the square and cube respectively while electron loss to the surface or through the outflow scales more or less proportionally to the plasma density. But such goes against the results  shown in Fig.\ \ref{fig:gainslosses_Mach}. Could this rather be due to a higher Mach number leading to higher temperatures which in turn increase the rate coefficients of the recombination processes? This can not be the reason because the rate coefficients either decrease with increasing temperature (in the case of 3-body recombination) or are weakly dependent on temperature (in the case of 2-body recombination).
\begin{figure*}[!t]
    \centering
    \subfigure[$P_{\rm dyn}=1$~kPa]{\includegraphics[width=0.35\textwidth]{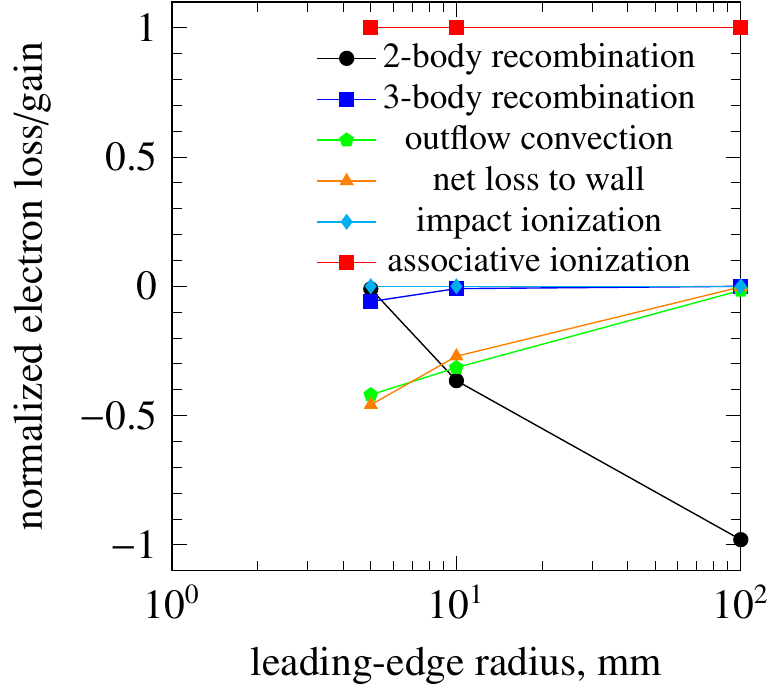}}
    \subfigure[$P_{\rm dyn}=10$~kPa]{\includegraphics[width=0.35\textwidth]{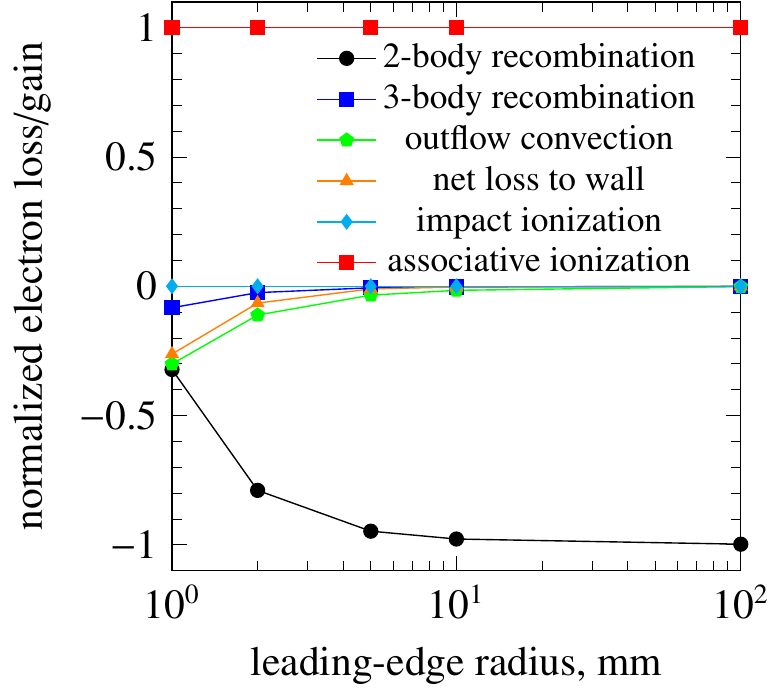}}
    \figurecaption{Impact of leading-edge radius on electron gains and losses at Mach 18 for (a) $P_{\rm dyn}=1$~kPa and (b) $P_{\rm dyn}=10$~kPa.}
    \label{fig:gainslosses_radius}
    %update results using Peclet
\end{figure*}

Rather, the recombination processes overtake other electron losses at high Mach number  because the plasma density increases by orders of magnitude with Mach number. This is due to most ions and electrons being created  through associative ionization (i.e., when atomic oxygen and nitrogen meet and form NO$^+$ and an electron). Thus, the density of the plasma depends strongly on the presence of the N and O radicals. Because the creation of these radicals goes up strongly as the translational temperature of the mixture is raised, and because the mixture temperature downstream of the bow shock reaches higher values for larger Mach numbers, an increase in flight Mach number leads to a much higher concentration of electrons and NO$^+$. Indeed, for all the cases shown in Fig.\ \ref{fig:gainslosses_Mach}, the electron density on the stagnation streamline becomes one hundred times higher or more as the Mach number is raised from 12 to 24. This explains why, as the flight Mach number is raised, the electron loss processes which scale with the square of the plasma density (i.e.\ 2-body recombination) or the cube of the plasma density (i.e.\ 3-body recombination) overtake electron loss processes which scale more or less linearly with plasma density (such as electron loss at the surface or through the outflow boundary).

\subsection{Effect of Leading Edge Radius}

We now proceed to study the impact of a two-order-of-magnitude change in leading edge radius  (from 1~mm to 10~cm) on electron losses. It is recalled that all dimensions on the wedge scale with the leading edge radius. Thus, a change in radius size can be thought of as a change in the scale of the vehicle leading edge. This parametric study is performed while keeping the flow Mach number fixed to 18 and setting the dynamic pressure either to 1 or to 10 kPa.

As shown in Fig.\ \ref{fig:gainslosses_radius}, at the highest radius considered of 10 cm, almost all electrons are lost through the 2-body dissociative recombination process. However, when the leading edge radius is in the order of millimeters, as many or more electrons are lost to the surface and to the outflow boundaries. The reason why the two-body recombination process loses some of its importance at small scales is because  the ratio between the volume of the flow behind the shock and the surfaces touching the flow becomes smaller as the scale is reduced. Indeed, when the freestream properties are not changed as is the case here, the amount of electrons lost through two-body dissociative recombination scales simply with the volume of the flow behind the shock because the reaction rates (which depend on densities and temperature) do not change much. On the other hand, the amount of electrons lost through reactions occurring on the surface is proportional to the surface area exposed to the plasma flow, not to the volume of plasma flow. Similarly, the amount of electrons lost through the outflow boundary is proportional to the area of the outflow boundary surface, not to the volume of plasma flow. As the scale of the vehicle is reduced, the ratio between the volume and the surface (of either the wall touching the plasma or the outflow boundary) becomes less. This leads to the two-body recombination losing some of its importance at small scales and getting overtaken by losses to the surface or through the outflow boundary.

\begin{figure}[!t]
    \centering
    \subfigure[$P_{\mathrm{dyn}}=3$~kPa]{\includegraphics[width=0.4\textwidth]{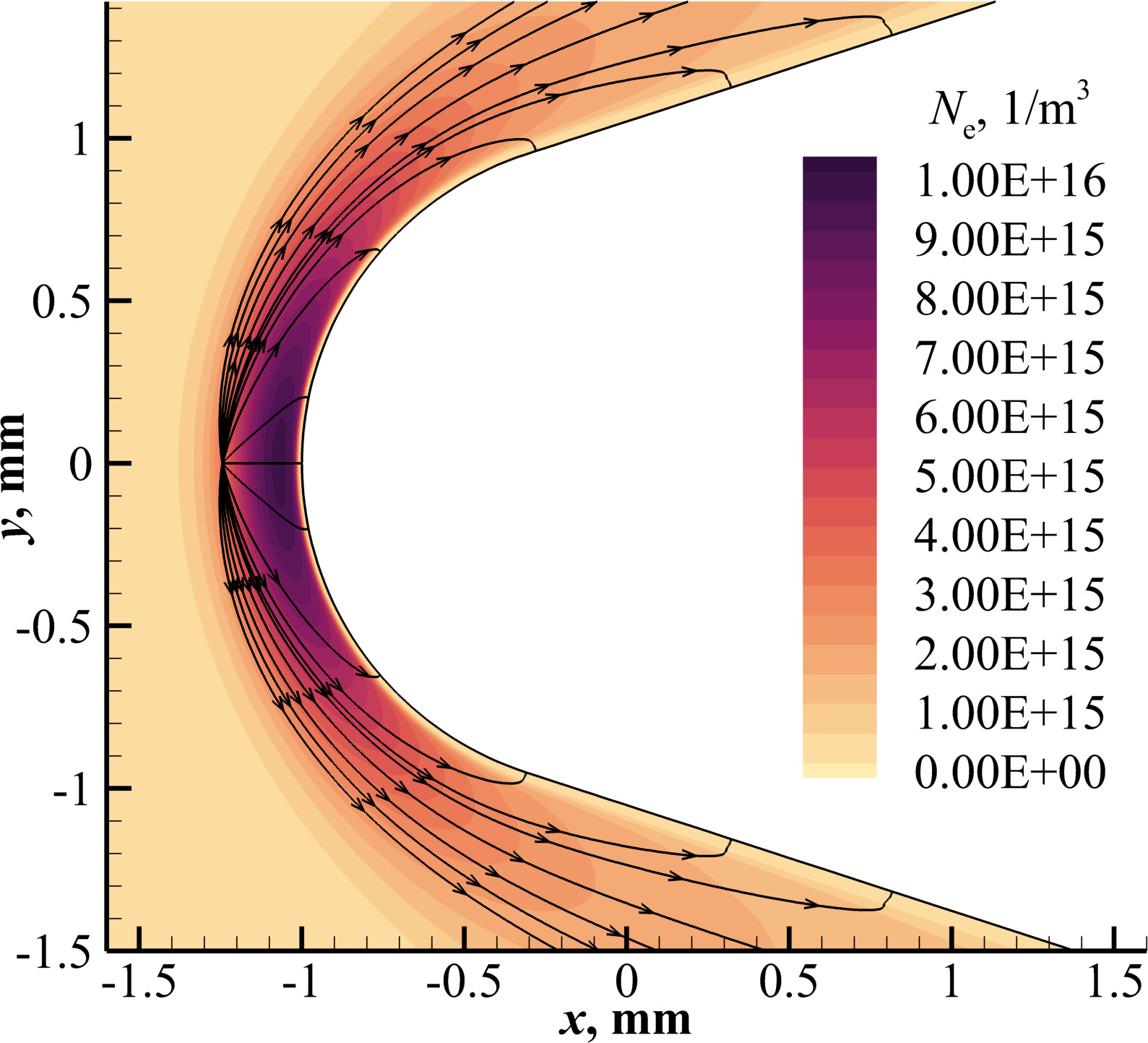}}
    \subfigure[$P_{\mathrm{dyn}}=50$~kPa]{\includegraphics[width=0.4\textwidth]{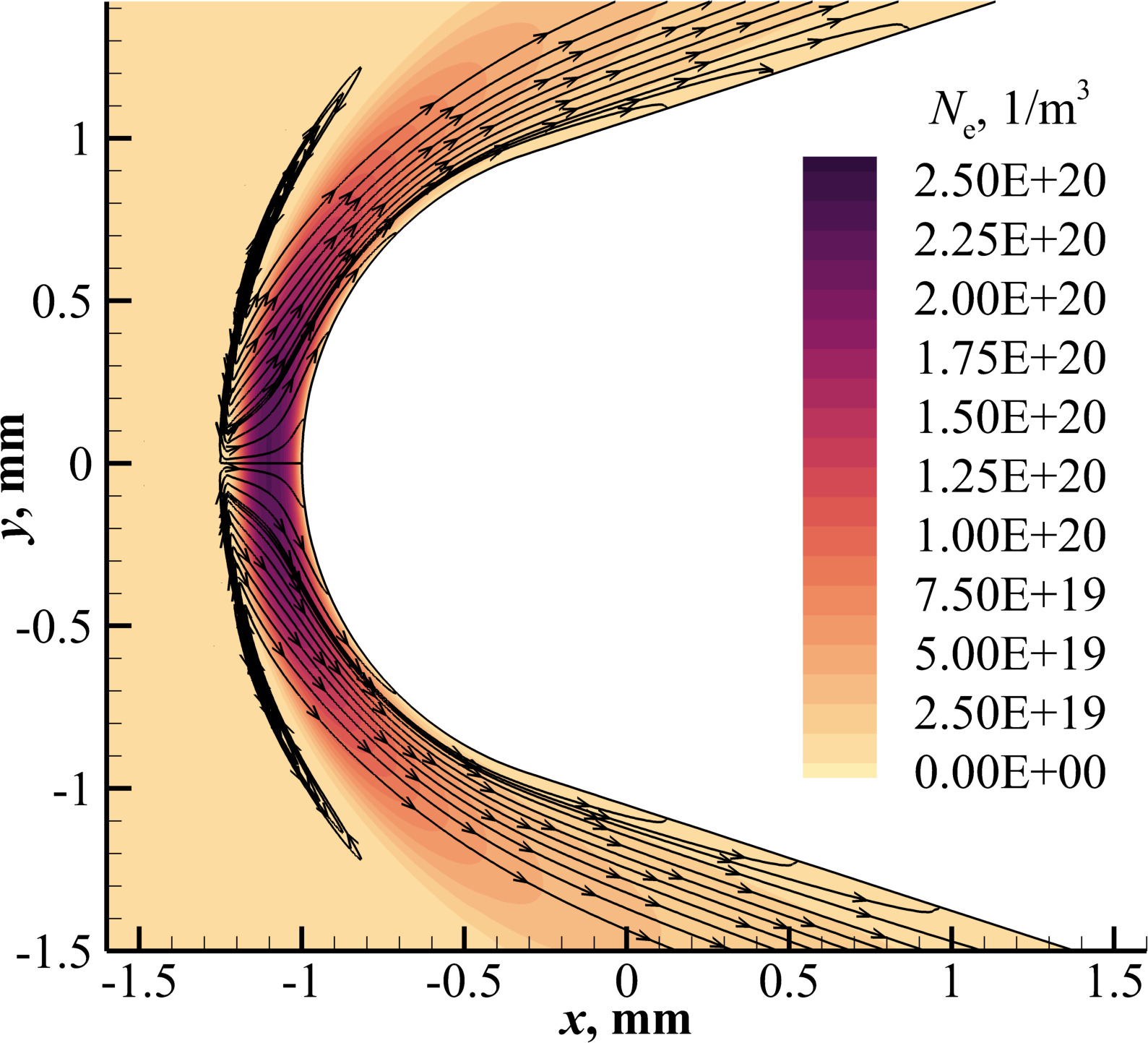}}
    \figurecaption{Electron velocity streamlines superimposed over electron number density contours  for $R=1$~mm, $M_\infty=18$, and (a) $P_{\mathrm{dyn}}=3$~kPa, (b) $P_{\mathrm{dyn}}=50$~kPa.}
    \label{fig:electronstreamlines}
\end{figure}

\subsection{Importance of the Plasma Sheath}

\begin{table*}[!h]
\centering
\scalefont{0.92}
\tablecaption{Comparison of maximum temperatures on the stagnation streamline and loss mechanisms for all cases. }
\begin{tabularx}{\linewidth}{*9{>{\centering\arraybackslash}X}}
\toprule
case number & flight dynamic pressure, kPa & Mach number & leading edge radius, mm & maximum $T$, K & maximum $T_{\rm v}$, K & maximum $T_{\rm e}$, K & \% of losses at surface & \% loss through 2-body recombination \\
\midrule
1 & 3 & 18 & 1 & 10911 & 1400 & 43 & 49 & 0 \\
2 & 10 & 18 & 1 & 11946 & 1578 & 835 & 26 & 32 \\
3 & 50 & 18 & 1 & 11888 & 3315 & 3008 & 1 & 92 \\
4 & 0.5 & 18 & 5 & 10604 & 1400 & 35 & 53 & 0 \\
5 & 1 & 18 & 5 & 11574 & 1400 & 145 & 46 & 1 \\
6 & 3 & 18 & 5 & 11970 & 1855 & 1425 & 12 & 67 \\
7 & 10 & 18 & 5 & 11885 & 3314 & 3028 & 1 & 95 \\
8 & 50 & 18 & 5 & 11578 & 5862 & 5338 & 0 & 99 \\
9 & 1 & 14 & 5 & 8212 & 1400 & 106 & 50 & 0 \\
10 & 1 & 16 & 5 & 9963 & 1400 & 102 & 48 & 0 \\
11 & 1 & 20 & 5 & 13048 & 1400 & 225 & 46 & 3 \\
12 & 1 & 22 & 5 & 14432 & 1400 & 314 & 45 & 4 \\
13 & 1 & 24 & 5 & 15779 & 1403 & 364 & 43 & 6 \\
14 & 3 & 14 & 5 & 8326 & 1635 & 829 & 37 & 14 \\
15 & 3 & 16 & 5 & 10222 & 1772 & 1189 & 21 & 49 \\
16 & 3 & 20 & 5 & 13611 & 1902 & 1552 & 9 & 74 \\
17 & 3 & 22 & 5 & 15234 & 1933 & 1623 & 7 & 77 \\
18 & 3 & 24 & 5 & 16925 & 1956 & 1668 & 6 & 77 \\
19 & 10 & 12 & 5 & 6375 & 2509 & 2210 & 32 & 24 \\
20 & 10 & 14 & 5 & 8282 & 2943 & 2642 & 6 & 82 \\
21 & 10 & 16 & 5 & 10150 & 3212 & 2885 & 2 & 93 \\
22 & 10 & 20 & 5 & 13546 & 3353 & 3130 & 0 & 95 \\
23 & 10 & 22 & 5 & 15198 & 3388 & 3230 & 0 & 95 \\
24 & 10 & 24 & 5 & 16947 & 3430 & 3326 & 0 & 95 \\
25 & 10 & 14 & 1 & 8347 & 1424 & 332 & 44 & 3 \\
26 & 10 & 16 & 1 & 10225 & 1513 & 614 & 36 & 15 \\
27 & 10 & 20 & 1 & 13547 & 1622 & 972 & 20 & 42 \\
28 & 10 & 22 & 1 & 15106 & 1651 & 1053 & 16 & 47 \\
29 & 10 & 24 & 1 & 16700 & 1672 & 1103 & 14 & 49 \\
30 & 1 & 18 & 10 & 11946 & 1578 & 856 & 27 & 37 \\
31 & 1 & 18 & 100 & 11783 & 4539 & 3793 & 0 & 98 \\
32 & 10 & 18 & 2 & 11956 & 2016 & 1833 & 7 & 79 \\
33 & 10 & 18 & 10 & 11776 & 4541 & 3788 & 0 & 98 \\
34 & 10 & 18 & 100 & 10142 & 5996 & 5977 & 0 & 100 \\
\bottomrule
\end{tabularx}
\label{tab:gainslosses_temperatures}
\end{table*}

The plasma sheath effects play a critical role in determining the losses. This is because the non-neutral sheath affects electron temperature through  electron cooling \cite{pf:2021:parent} and electron temperature affects recombination rates as well as ambipolar diffusion to the surface.

How can the relatively thin sheath lead to significant electron cooling of the much larger plasma bulk? Electron cooling occurs in the sheath because electromagnetic energy input in the sheath is strongly negative.  Indeed, the electron electromagnetic energy input can be expressed as $\vec{E}\cdot\vec{J}_{\rm e}$ with $\vec{E}$ the electric field and $\vec{J}_{\rm e}$ the electron current. Within a sheath that borders a dielectric (as is the case here all around the wedge), there is no net current to the surface. Thus, the electron current coming out of the surface must balance the ion current going towards the surface. Further, the electric field points towards the surface while the electron current points in the opposite direction. Because the electric field is high and the electron current due to diffusion is also high, this leads to $\vec{E}\cdot\vec{J}_{\rm e}$ being negative and having a large magnitude thus resulting in cooling of the electrons. Although the thickness of the non-neutral sheath typically does not exceed one fifth of the distance between the body and the bow shock (see electron streamlines in Fig.\ \ref{fig:electronstreamlines} from which the sheath thickness can be deduced), the cooling is not limited to the sheath region and spreads to the rest of the plasma flow through the fast electron thermal conductivity process. Thus, the lowering of the electron temperature due to sheath cooling occurs not only within the non-neutral sheath near the surface but also within the quasi-neutral plasma flow far from the surfaces. 

The amount of electron cooling is here assessed by monitoring the maximum electron temperature on the stagnation streamline, as shown in Table \ref{tab:gainslosses_temperatures}. An interesting relationship can be observed between electron loss to the surface and electron cooling.  For cases where there is a large amount of electron cooling, the maximum electron temperature on the stagnation streamline drops to below-freestream values (i.e.\, less than 240~K). Every time this occurs, the electron losses to the surface strongly dominate over the recombination losses through the plasma. On the other hand, for cases with a relatively small amount of electron cooling, the maximum electron temperature on the stagnation streamline reaches high values approaching the maximum vibrational temperature. When this occurs, the dominant electron loss mechanism changes to 2-body recombination within the plasma. 

It is emphasized that, although electron loss through surface catalyticity becomes dominant only when electron cooling is significant, this does not mean that there is a causal relationship at play. In fact, we see here clearly that one effect does not cause the other. If electron cooling would cause surface losses to become more important, then we would see an increase in ambipolar diffusion due to cooling. But the opposite is rather observed: more cooling leads to a reduced electron temperature and this in turn leads to a smaller ambipolar diffusion coefficient $(1+T_{\rm e}/T)$. 

Electron cooling does not only affect surface catalyticity but also affects  2-body dissociative recombination processes within the plasma bulk because the latter also depends on electron temperature. Both processes are affected differently, however. While a decrease in electron temperature leads to less electron loss to the surface, it rather leads to more losses within the plasma bulk. Thus, for cases where the amount of electrons lost at the surface is about the same as within the plasma bulk, the two effects may cancel each other more or less and the electron cooling by the sheath does not result in a significant net change in plasma density. But such a cancellation of the effects does not always occur. Indeed, when most of the electrons are lost at the surface and few are lost within the plasma bulk, then the cooling of electrons by the sheath leads to a significant difference in plasma density. In other words, if the electron cooling by the sheaths would not be included in the physical model, the numerical simulations would overpredict considerably electron loss to the surface and underpredict plasma density especially when electron loss to the surface is the dominant loss mechanism.

\subsection{Limitations of the Physical Model}

{\color{red}
Because the physical model relies on continuum mechanics through the use of fluid transport equations, it would lead to significant error when the mean free path is commensurate with the distance between the shock and the surface. It has been verified that this is not a source of concern except for the cases that have both (i)  the lowest freestream dynamic pressure considered of 1~kPa and (ii) the lowest leading edge radius considered of 1~mm. For these cases, there are only a few collisions between the shock edge and the sheath and the continuum approximation is at the limit of being valid. Therefore, for these extreme cases, a more accurate representation of the flow physics may be obtained through the use of physical models that track collisions between individual molecules such as those used within DSMC (Direct Simulation Monte Carlo) \cite{book:1994:bird,book:2017:boyd}.

The capability of the drift-diffusion fluid model  in capturing the sheath physics accurately has also been verified by comparing the sheath thickness to the mean free path of the molecules. It was found that for all cases here considered, there were between 2 and 5 collisions within the sheath. Sheath models that are collisionless would hence not be valid here, and our choice of a collisional sheath model through the drift-diffusion approximation is justified. Nonetheless, a more accurate representation of the sheath physics could be obtained through kinetic simulations that include collisions. One advantage that the kinetic model has over the drift-diffusion model is that it does not lump all the particules into one group with one common average temperature. Rather, the electrons are separated into several groups each with its own energy level. This would permit to simulate more accurately the phenomenon of electron cooling due to highly energetic electrons going through the sheath  and getting absorbed by the surface. Such a phenomenon is taken into consideration into the electron energy transport equation used herein through the $\vec{E} \cdot \vec{J}_{\rm e}$ term being negative within the sheath. But such a model for electron cooling can lead to error because it lumps both the low energy electrons and high energy electrons together. 

Simulating hypersonic non-neutral plasma flows using a combination of kinetic simulations for the sheath and DSMC for the shock and plasma layer is non trivial, however. As seen in this paper the sheath and the plasma bulk can not be simulated independently of each other because of the tight coupling between the two regions. Due to this coupling, numerical difficulties will arise when combining one physical model for the sheath and another different physical model for the plasma bulk. Perhaps a better avenue would be to calibrate, through comparisons with DSMC and kinetic simulations, the transport coefficients used in the fluid models outlined herein to make them more accurate at very low flight dynamic pressures and leading edge radii.}

\section{Conclusions}
%\appendix
%\section{Appendix Test}   

Using advanced numerical methods, the first comprehensive study of electron gains and losses in hypersonic air flow around a wedge in the Mach number range 12--24 and dynamic pressure range 1--50~kPa is here conducted. The physical model includes an improved  11-species Park chemical solver where all the reaction rates involving electrons are adjusted using BOLSIG+ and recent cross-sectional data. Another aspect of the physical model that is an improvement over previous work is the inclusion of the non-neutral plasma sheaths and their coupling to the quasi-neutral plasma bulk.

For the range of flight conditions considered, electron gains are found to be due almost entirely to associative ionization of nitric oxide. Other ionization mechanisms such as electron impact ionization or associative ionization of oxygen or nitrogen are seen to play a negligible role.  As for electron losses, the dominant loss mechanism is seen to vary significantly depending on the freestream conditions and the scale of the vehicle. Electron loss to the surface through catalyticity dominates over electron-ion recombination within the flow either (i) at high altitude where the dynamic pressure is low, (ii) at low Mach number, or (iii) when the vehicle has a sharp leading edge. 

Including the non-neutral plasma sheaths within the physical model is seen to be critical to predict accurately plasma density especially when the dominant electron loss mechanism is surface catalyticity. This is because  (i) whenever electron loss to the surface is the dominant loss mechanism, we observe a large amount of electron cooling coming from the non-neutral sheaths which affects significantly electron temperature everywhere in the plasma, and  (ii) an accurate determination of electron temperature in the plasma bulk is critical to assess losses to the surface because such are mostly function of ambipolar diffusion which depends on electron temperature.

\section{Acknowledgments}
This work has been funded in part by Raytheon Missiles and Defense.  The First Author would like to acknowledge discussions with Sergey Macheret, Mikhail Shneider, Kyle Hanquist, Kurt Elkins, Mark Meisner, and Brett Ridgely. 

\section{Data Availability}

The data is available on request from the authors.

\bibliography{all}
\bibliographystyle{aiaa}

\end{document}